\documentclass[useAMS,usenatbib]{mn2e}
\usepackage{graphicx}
\usepackage{lscape}

\newcommand{\Lsun}{\mbox{$L_{\odot}$}} 
\newcommand{\ltsimeq}{\raisebox{-0.6ex}{$\,\stackrel
        {\raisebox{-.2ex}{$\textstyle <$}}{\sim}\,$}}

\title[Very cool T dwarfs]{Exploring the substellar temperature regime
  down to $\sim$550K}
\author[Ben Burningham et al.]{Ben Burningham$^{1}$\thanks{E-mail: B.Burningham@herts.ac.uk}, 
D. J. Pinfield$^{1}$, S. K. Leggett$^{2}$, M. Tamura$^{3}$, P. W. Lucas$^{1}$, 
\newauthor
 D. Homeier$^{4}$, A. Day-Jones$^{1}$,
 H. R. A. Jones$^{1}$, J.R.A. Clarke$^{1}$,
\newauthor
 M. Ishii$^{5}$, M. Kuzuhara$^{6}$, N. Lodieu$^{7}$, M. R. Zapatero Osorio$^{7}$,
 B.P. Venemans$^{8}$,  
\newauthor
D.J. Mortlock$^{9}$, D. Barrado y Navascu\'es$^{10}$,E. L. Martin$^{7}$, A. Magazz\`u$^{11}$ \\  
$^{1}$Centre for Astrophysics Research, Science and Technology Research
Institute, University of Hertfordshire, Hatfield AL10 9AB \\
$^{2}$Gemini Observatory, 670 N. A'ohoku Place, Hilo, HI 96720, USA \\
$^{3}$National Astronomical Observatory, Mitaka, Tokyo 181-8588 \\
$^{4}$Institut fur Astrophysik, Georg-August-Universitat, Friedrich-Hund-Platz 1,
     37077 Gottingen, Germany \\
$^{5}$Subaru Telescope, 650 North A'ohoku Place, Hilo, Hi 96720, USA \\
$^{6}$University of Tokyo, Hongo, Tokyo 113-0033, Japan\\
$^{7}$Instituto de Astrof\'isica de Canarias, 38200 La Laguna, Spain \\
$^{8}$Institute of Astronomy, Madingley Road, Cambridge CB3 0HA, UK \\
$^{9}$Astrophysics Group, Imperial College London, Blackett Laboratory,
     Prince Consort Road, London SW7 2AZ \\
$^{10}$Laboratorio de Astrof\'isica Espacial y F\'isica
     Fundamental,INTA, P.O. Box 78, E--28691 Villanueva de la Canada (Madrid), Spain \\
$^{11}$Fundaci\'on Galileo Galilei-INAF, Apartado 565, E-38700
        Santa Cruz de La Palma, Spain 
}

\begin{document}
%
%  These Macros are taken from the AAS TeX macro package version 4.0.
%  Include this file in your LaTeX source only if you are not using
%  the AAS TeX macro package and need to resolve the macro definitions
%  in the BibTeX entries returned by the ADS abstract service.
%
%  If you plan not to use this file to resolve the journal macros
%  rather than the whole AAS TeX macro package, you should save the
%  file as ``aas_macros.sty'' and then include it in your paper by
%  using a construct such as:
%	\documentstyle[11pt,aas_macros]{article}
%
%  For more information on the AASTeX macro package, please see the URL
%	http://www.aas.org/publications/aastex.html
%  For more information about ADS abstract server, please see the URL
%	http://adswww.harvard.edu/ads_abstracts.html
%

% Abbreviations for journals.  The object here is to provide authors
% with convenient shorthands for the most "popular" (often-cited)
% journals; the author can use these markup tags without being concerned
% about the exact form of the journal abbreviation, or its formatting.
% It is up to the keeper of the macros to make sure the macros expand
% to the proper text.  If macro package writers agree to all use the
% same TeX command name, authors only have to remember one thing, and
% the style file will take care of editorial preferences.  This also
% applies when a single journal decides to revamp its abbreviating
% scheme, as happened with the ApJ (Abt 1991).

\def\aj{\rm{AJ}}                   % Astronomical Journal
\def\araa{\rm{ARA\&A}}             % Annual Review of Astron and Astrophys
\def\apj{\rm{ApJ}}                 % Astrophysical Journal
\def\apjl{\rm{ApJ}}                % Astrophysical Journal, Letters
\def\apjs{\rm{ApJS}}               % Astrophysical Journal, Supplement
\def\ao{\rm{Appl.~Opt.}}           % Applied Optics
\def\apss{\rm{Ap\&SS}}             % Astrophysics and Space Science
\def\aap{\rm{A\&A}}                % Astronomy and Astrophysics
\def\aapr{\rm{A\&A~Rev.}}          % Astronomy and Astrophysics Reviews
\def\aaps{\rm{A\&AS}}              % Astronomy and Astrophysics, Supplement
\def\azh{\rm{AZh}}                 % Astronomicheskii Zhurnal
\def\baas{\rm{BAAS}}               % Bulletin of the AAS
\def\jrasc{\rm{JRASC}}             % Journal of the RAS of Canada
\def\memras{\rm{MmRAS}}            % Memoirs of the RAS
\def\mnras{\rm{MNRAS}}             % Monthly Notices of the RAS
\def\pra{\rm{Phys.~Rev.~A}}        % Physical Review A: General Physics
\def\prb{\rm{Phys.~Rev.~B}}        % Physical Review B: Solid State
\def\prc{\rm{Phys.~Rev.~C}}        % Physical Review C
\def\prd{\rm{Phys.~Rev.~D}}        % Physical Review D
\def\pre{\rm{Phys.~Rev.~E}}        % Physical Review E
\def\prl{\rm{Phys.~Rev.~Lett.}}    % Physical Review Letters
\def\pasp{\rm{PASP}}               % Publications of the ASP
\def\pasj{\rm{PASJ}}               % Publications of the ASJ
\def\qjras{\rm{QJRAS}}             % Quarterly Journal of the RAS
\def\skytel{\rm{S\&T}}             % Sky and Telescope
\def\solphys{\rm{Sol.~Phys.}}      % Solar Physics
\def\sovast{\rm{Soviet~Ast.}}      % Soviet Astronomy
\def\ssr{\rm{Space~Sci.~Rev.}}     % Space Science Reviews
\def\zap{\rm{ZAp}}                 % Zeitschrift fuer Astrophysik
\def\nat{\rm{Nature}}              % Nature
\def\iaucirc{\rm{IAU~Circ.}}       % IAU Cirulars
\def\aplett{\rm{Astrophys.~Lett.}} % Astrophysics Letters
\def\apspr{\rm{Astrophys.~Space~Phys.~Res.}}
                % Astrophysics Space Physics Research
\def\bain{\rm{Bull.~Astron.~Inst.~Netherlands}} 
                % Bulletin Astronomical Institute of the Netherlands
\def\fcp{\rm{Fund.~Cosmic~Phys.}}  % Fundamental Cosmic Physics
\def\gca{\rm{Geochim.~Cosmochim.~Acta}}   % Geochimica Cosmochimica Acta
\def\grl{\rm{Geophys.~Res.~Lett.}} % Geophysics Research Letters
\def\jcp{\rm{J.~Chem.~Phys.}}      % Journal of Chemical Physics
\def\jgr{\rm{J.~Geophys.~Res.}}    % Journal of Geophysics Research
\def\jqsrt{\rm{J.~Quant.~Spec.~Radiat.~Transf.}}
                % Journal of Quantitiative Spectroscopy and Radiative Transfer
\def\memsai{\rm{Mem.~Soc.~Astron.~Italiana}}
                % Mem. Societa Astronomica Italiana
\def\nphysa{\rm{Nucl.~Phys.~A}}   % Nuclear Physics A
\def\physrep{\rm{Phys.~Rep.}}   % Physics Reports
\def\physscr{\rm{Phys.~Scr}}   % Physica Scripta
\def\planss{\rm{Planet.~Space~Sci.}}   % Planetary Space Science
\def\procspie{\rm{Proc.~SPIE}}   % Proceedings of the SPIE

\let\astap=\aap
\let\apjlett=\apjl
\let\apjsupp=\apjs
\let\applopt=\ao

\maketitle

\begin{abstract}

We report the discovery of three very late T~dwarfs in the UKIRT Infrared
Deep Sky Survey (UKIDSS) Third Data Release: ULAS~J101721.40+011817.9
(ULAS1017), ULAS~J123828.51+095351.3 (ULAS1238) and
ULAS~J133553.45+113005.2 (ULAS1335).
We detail optical and near-infrared photometry for all three sources,
and mid-infrared photometry for ULAS1335.
We use near-infrared spectra of each source to assign spectral types
T8p (ULAS1017), T8.5 (ULAS1238) and T9 (ULAS1335) to these objects. 
ULAS1017 is classed as a peculiar T8 (T8p) due to appearing as a
T8~dwarf in the $J$-band, whilst exhibiting $H$~and $K$-band flux
ratios consistent with a T6 classification.
Through comparison to BT-Settl model spectra we estimate that ULAS1017 has $750 {\rm K} \ltsimeq T_{\rm eff}
\ltsimeq 850 {\rm K}$, and $5.0 \ltsimeq \log {\it g }(cms^{-2}) \ltsimeq  5.5$,
assuming solar metallicity. This estimate for gravity is degenerate
with varying metallicity.
We estimate that ULAS1017 has an age of 1.6--15~Gyr, a mass of
33--70~M$_J$ and lies at a distance of 31--54~pc.
We do not estimate atmospheric parameters for ULAS1238 due to a lack
of $K$-band photometry.
We extend the unified scheme of \citet{burgasser06} to the type T9
and suggest the inclusion of the $W_J$ index to replace the now
saturated $J$-band indices. We propose ULAS1335 as the T9 spectral
type standard. 
ULAS1335 is the same spectral type as ULAS~J003402.77-005206.7 and
CFBDS~J005910.90-011401.3. 
We argue that given the similarity of the currently known
$>$T8~dwarfs to the rest of the T~dwarf sequence, the suggestion of
the Y0 spectral class for these objects is premature.
Comparison of model spectra with that of ULAS1335 suggest a
temperature below 600K, possibly combined with low-gravity and/or
high-metallicity.
We find ULAS1335 to be extremely red  in near to mid-infrared colours, with
$H -[4.49]~=~4.34 \pm 0.04$ . This is the reddest near to mid-infrared colour
yet observed for a T~dwarf.  
The near to mid-infrared spectral energy distribution of ULAS1335
further supports $T_{\rm eff} < 600{\rm K}$, and we estimate $T_{\rm eff}
  \sim 550-600{\rm K}$ for ULAS1335.
We estimate that ULAS1335 has an age of 0.6--5.3~Gyr, a mass of
15--31~M$_J$ and lies at a distance of 8--12~pc.

\end{abstract}


\section{Introduction}
\label{sec:intro}
Extending the known sample of field dwarfs to ever lower effective temperatures
 ($T_{\rm eff}$) 
 is important not only for the determination of the field
mass function, but also for probing temperature and pressure regimes
that have hitherto been unexplored observationally. 
The study of extremely cool brown dwarfs opens a window on
atmospheric physics which will be fundamental for understanding
the processes within a broad range of substellar atmospheres,
including those of giant exoplanets.

At the time of writing there are 142 T~dwarfs published in the
literature \citep[e.g., DwarfArchives.org; ][]{pinfield08}. 
Of these, only five are
classified as T8 or later \citep[using the scheme of][]{burgasser06},
two of which have spectral types later than T8:
ULAS~J003402.77-005206.7 \citep[hereafter ULAS0034; ][]{warren07} and
CFBDS~J005910.90-011401.3 \citep[hereafter CFBDS0059; ][]{delorme08}.
ULAS0034 has an inferred $T_{ \rm eff}$ of 600-700~K \citep{warren07}, and 
\citet{delorme08} infer a $T_{ \rm eff}$ for CFBDS0059 that is $\sim 50$~K
cooler than that of ULAS0034.

Expanding the sample of very cool brown dwarfs, and exploring
possible factors that could motivate the implementation of new
spectral classes \citep[e.g Y dwarfs; ][]{kirkpatrick99} are primary
science drivers for the UKIRT 
Infrared Deep Sky Survey (UKIDSS) Large Area Survey \citep[LAS; see][]{ukidss}.
Our collaboration, the UKIDSS Cool Dwarf Science Working Group (CDSWG)
is engaged in a substantial effort to achieve this aim
\citep[e.g.,][]{pinfield08,kendall07,lod07,warren07}. 
Here we present the recent discovery by CDSWG of three very cool brown
dwarfs. 
These are ULAS~J101721.40+011817.9 (ULAS1017),
ULAS~J123828.51+095351.3 (ULAS1238) and ULAS~J133553.45+113005.2
(ULAS1335).
The last of these, ULAS1335, may be the coolest brown dwarf yet
discovered and we here discuss the properties of this object,
estimating a $T_{\rm eff}$ of approximately 550--600~K.

\section{Source identification}
\label{sec:ident}
The objects presented here were selected from the third data release (DR3)
of UKIDSS (Warren et al., in prep).
%\citep{warrendr3}. 
We used the same method as that outlined in Sections~2 and~3 of
 \citet{pinfield08}. 
 The process of identifying and confirming extremely late T~dwarfs can
be separated into three principle stages: 1) initial selection by
mining survey data; 2) confirmation of photometric properties by
near-infrared and optical follow-up; 3) near-infrared spectroscopy to
estimate spectral type. 
The Gemini/NIRI spectroscopic follow-up took place in two stages in
queue mode. 
Initial, short ($t_{int}= 16$mins) $J$-band observations were obtained
first. 
These initial spectra confirmed that two of our targets were indeed
of extremely late spectral type, and deeper $J$, $H$~and $K$-band
observations were then queued.
Subaru/IRCS spectroscopic follow-up took place
separately for additional candidates in classical mode and confirmed the very
late type of a third target.
We will discuss each of these stages in more
detail in the following sections.

\subsection{Mining UKIDSS DR3}
\label{subsec:mining}

We initially searched the UKIDSS LAS data for sources that met colour criteria:
$Y-J > 0.5$; $J-H < 0.1$; $\sigma < 0.3$, where $\sigma$ is the
uncertainty in the magnitude measured in each of the three
$YJH$-bands.
The resultant candidate list was cross-matched against the Sloan
Digital Sky Survey (SDSS) Data Release 6 (DR6) using a pairing radius
of 2 arcsecs. 
Objects with no SDSS counterpart, or whose SDSS-UKIDSS $z'-J > 3.0$, were retained as
potential candidates.
We identified some 50 candidate T~dwarfs in this search.
ULASJ1238 and ULASJ1335 were picked out for immediate and rapid
follow-up due to survey colours that were reminiscent of the
prototype $>$T8 dwarf ULAS J0034.
ULAS1017 on the other hand was selected due to its extremely blue
$J-H$ colour from UKIDSS photometry.

\subsection{Ground based photometric follow-up}
\label{subsec:photo}

Near infrared follow-up photometry was obtained using the UKIRT Fast
Track Imager \citep[UFTI;][]{roche03} mounted on UKIRT, and the
Long-slit Infrared Imaging Spectrograph \citep[LIRIS;][]{Manchado98}
mounted on the William Herschel Telescope on La Palma.
The mosaics were produced using sets of jittered images, with
individual exposure times, jitter patterns and number of repeats given
in Table~\ref{tab:photobs}.
The data were dark subtracted, flatfield corrected, sky subtracted and
mosaiced using ORAC-DR for the UFTI data, and LIRIS-DR for the LIRIS data. 
We calibrated our UFTI observations using
UKIRT Faint Standards \citep{ukirtfs}, with a standard observed at a similar airmass
for each target. 
All UFTI data were obtained under photometric conditions, with seeing
better than 0.9 arcsec. 
The wider field LIRIS data were obtained through thin cirrus in variable seeing
(0.8 - 2.0 arcsec) and were calibrated against 2MASS stars within the
field of view. 
LIRIS uses a $K_s$ filter, and as such no transform was required for
the 2MASS stars. 
The $K_s$ magnitude for ULAS1017 was transformed to the MKO
system using the transform derived by \citet{pinfield08}. 

We have also obtained optical $z$-band photometry using the ESO
Multi-Mode Instrument (EMMI) mounted on the New Technology Telescope
at La Silla, Chile on the nights of 2008 January 29 and 2008
January 30.
These observations are summarised in Table~\ref{tab:photobs}.  
For this optical follow-up we used a Bessel $z$-band filter (ESO Z\#611). 
The data were reduced using standard IRAF packages, and then multiple
images of the same target were aligned and stacked to increase signal-to-noise.

We calibrated our $z-$band photometry using SDSS DR6 sources present
in our images as secondary standards. 
We transformed the Sloan $z'(AB)$ magnitudes to the EMMI system
using the transformation given by \citet{warren07} before using them
to determine zero points, which had a typical scatter of $\pm 0.04$ mags.
The resulting EMMI photometry for our targets was then transformed to
the Sloan $z'(AB)$ system \citep[see][]{warren07}.
The results of our ground-based follow-up photometry are
given in Table~\ref{tab:mags}, which also gives the original database
photometry.

 \begin{table*}
\begin{tabular}{| c | c c c c c |}
  \hline
Object & Filter & Instrument & UT Date & Total integration time & $t_{\rm int}$ breakdown  \\
\hline
 ULAS1017  & $J$ & UFTI & 2008 Jan 17 & 300s & (j = 5, r = 1, $t_{\rm exp}$ = 60s)\\
           & $H$ & UFTI & 2008 Jan 17 & 2400s & (j = 5, r = 6, $t_{\rm exp}$ = 60s)\\
           & $K_{\it s}$ & LIRIS & 2008 Mar 17 & 2400s & (j = 5, r = 24, $t_{\rm exp}$ = 20s) \\
           & $z_{\rm EMMI}$ & EMMI & 2008 Jan 30 & 2400s & (j = 1,r = 4, $t_{\rm exp}$ = 600s)\\
 ULAS1238  & $J$ & UFTI & 2008 Jan 25 & 300s & (j = 5, r = 1, $t_{\rm exp}$ = 60s)\\
             & $H$ & UFTI & 2008 Jan 25 & 1800s & (j = 5, r = 6, $t_{\rm exp}$ = 60s)\\
             & $z_{\rm EMMI}$ & EMMI & 2008 Jan 30 & 2400s & (j = 1,r = 4, $t_{\rm exp}$ = 600s)\\
 ULAS1335  & $Y$ & UFTI & 2008 Jan 16 & 540s & (j = 9, r = 1, $t_{\rm exp}$ = 60s)    \\
             & $J$ & UFTI & 2008 Jan 16 & 300s & (j = 5, r = 1, $t_{\rm exp}$ = 60s) \\
             & $H$ & UFTI & 2008 Jan 16 & 900s & (j = 5, r = 3, $t_{\rm exp}$ = 60s) \\
             & $K$ & UFTI & 2008 Jan 16 & 900s & (j = 5, r = 3, $t_{\rm exp}$ = 60s) \\
             & $z_{\rm EMMI}$ & EMMI & 2008 Jan 29 & 1200s & (j = 1,r = 2, $t_{\rm exp}$ = 600s) \\
\hline
\end{tabular}
\caption{Summary of the observations obtained for near infrared
  and optical photometric follow-up. The breakdown of each integration
  is given in the final column with the following notation: j = number
  of jitter points; r = number of repeats for jitter pattern;
  $t_{\rm exp}$ = exposure time at each jitter point.
\label{tab:photobs}}

\end{table*}

\begin{table*}
\begin{tabular}{| c | c c c c c c c c c c c c c |}
  \hline
Object & $Y$ & $J$ & $H$ & $K$ & $z'(AB)$ & $z'(AB) - J$ & $Y-J$ & $J-H$ & $H-K$    \\
\hline
 ULAS1017  & $19.72 \pm 0.15$ &  $18.57 \pm 0.07$  & $19.38 \pm 0.24$  &  &  &  & $1.15 \pm 0.17$  & $-0.82 \pm 0.25$  & \\
           &           &  $18.53 \pm 0.02$  & $19.07 \pm 0.02$   & $19.01 \pm 0.10$   & $22.49 \pm 0.09$  & $3.96 \pm 0.09$  & $1.19 \pm 0.15$  & $-0.54 \pm 0.03$  & $0.06 \pm 0.10$ \\
 ULAS1238  & $19.55 \pm 0.07$  & $18.78 \pm 0.06$  & $19.25 \pm 0.15$  &  &  &  & $0.77 \pm 0.09$  & -$0.47 \pm 0.17$  &  \\
             &           & $18.95 \pm 0.02$  & $19.20 \pm 0.02$   &  & $22.74 \pm 0.12$  & $3.79 \pm 0.12$   & $0.60 \pm 0.07$  & $-0.25 \pm 0.03$  & \\
 ULAS1335  & $18.70 \pm 0.05$  & $17.90 \pm 0.03$  & $18.15 \pm 0.12$  & $18.47 \pm 0.20$  & & & $0.80 \pm 0.06$  & -$0.25 \pm 0.13$  & -$0.32 \pm 0.23$  \\ 
             & $18.81 \pm 0.04$  & $17.90 \pm 0.01$  & $18.25 \pm 0.01$   & $18.28 \pm 0.03$  &  $22.04 \pm 0.10$  & $4.14 \pm 0.10$  & $0.91 \pm 0.04$  & -$0.35 \pm 0.01$  & -$0.03 \pm 0.03$  \\

\hline
\end{tabular}
\caption{The available photometry for each of the new ultracool dwarfs. In
  each case the first row gives survey photometry from UKIDSS LAS
  \citep[apermag3; see][]{dye06}.  All three objects were undetected in
  SDSS DR6. The second row for each object gives the results of the
  follow-up photometry described in Section~\ref{subsec:photo}.
\label{tab:mags}
}

\end{table*}

\subsection{Spectroscopic confirmation}
\label{subsec:spectra}

Spectroscopy in the $JHK$-bands was obtained for ULAS1238 and
ULAS1335 using the Near InfraRed
Imager and Spectrometer \citep[NIRI;][]{hodapp03} on the Gemini North
Telescope on Mauna Kea under programs GN-2007B-Q-26 and GN-2008A-Q-15. 
The InfraRed Camera and Spectrograph \citep[IRCS;][]{IRCS2000} on the
Subaru telescope on Mauna Kea was used to obtain the $JH$ spectrum for
ULAS1017 and this was then followed with spectroscopy in the $H$ and 
$K$-bands with Gemini/NIRI. 
Additionally, we obtained a $Y$-band spectrum for ULAS1335 using
the Infrared Spectrometer And Array Camera \citep[ISAAC; ][]{isaac}
mounted on VLT UT1 and ESO, Paranal in Director's Discretionary time
(program ID: 280.C-5067(A)).

All observations were made up of a set of sub-exposures in an ABBA
jitter pattern to facilitate effective background subtraction, with a
slit width of 1 arcsec. 
The length of the A-B jitter was 10 arcsecs.
 The observations are summarised in Table~\ref{tab:specobs}.
The NIRI observations were reduced using standard IRAF
Gemini packages. 
The Subaru IRCS $JH$ spectrum was also extracted using standard IRAF
packages. The AB pairs were subtracted using generic IRAF tools,
and median stacked.  
In the case of IRCS, the data were found to be sufficiently uniform in
the spatial axis for flat-fielding to be neglected.
The ISAAC data were reduced and extracted using the same steps, but
implemented with the ESO ISAAC pipeline
version~5.7.0.

For all three sets of data, a comparison argon arc frame was
used to obtain a dispersion solution, which was then applied to the
pixel coordinates in the dispersion direction on the images.
The resulting wavelength-calibrated subtracted pairs had a low-level
of residual sky emission removed by fitting and subtracting this
emission with a set of polynomial functions fit to each pixel row
perpendicular to the dispersion direction, and considering pixel data
on either side of the target spectrum only. 
The spectra were then extracted using a linear aperture, and cosmic
rays and bad pixels removed using a sigma-clipping algorithm.

Telluric correction was achieved by dividing each extracted target
spectrum by that of an early A or F type star observed just before or
after the target and at a similar airmass.
Prior to division, hydrogen lines were removed from the standard star
spectrum by interpolating the stellar
continuum.
Relative flux calibration was then achieved by multiplying through by a
blackbody spectrum of the appropriate $T_{\rm eff}$.
Data obtained for the same spectral regions on different nights were
co-added after relative flux calibration, each weighted by their exposure time.

 \begin{table*}
\begin{tabular}{| c | c c c c c |}
  \hline
Object & UT Date & Integration time & Instrument &
Spectral region \\
\hline
  ULAS1017  & 2008 Jan 25 & 12x300s & IRCS & $JH$ \\
            & 2008 Mar 14 & 16x225s & NIRI & $K$ \\
            & 2008 May 11 & 12x300s & NIRI & $H$ \\
  ULAS1238  & 2007 Dec 21 & 4x240s & NIRI & $J$ \\
            & 2008 Jan 23 & 12x300s & NIRI & $J$ \\
            & 2008 Jan 23 & 12x300s & NIRI & $H$ \\
            & 2008 Feb 18 & 8x225s & NIRI & $K$ \\
            & 2008 Feb 26 & 6x225s & NIRI & $K$ \\
  ULAS1335  & 2007 Dec 21 & 4x240s & NIRI & $J$ \\
            & 2008 Jan 22 & 12x300s & NIRI & $J$ \\
            & 2008 Jan 22 & 12x300s & NIRI & $H$ \\
            & 2008 Feb 28 & 20x225s & NIRI & $K$ \\
            & 2008 Mar 19 & 12x300s & ISAAC & $Y$ \\
\hline
\end{tabular}
\caption{Summary of the near-infrared spectroscopic observations.
\label{tab:specobs}}

\end{table*}

\begin{figure}
\includegraphics[height=250pt, angle=90]{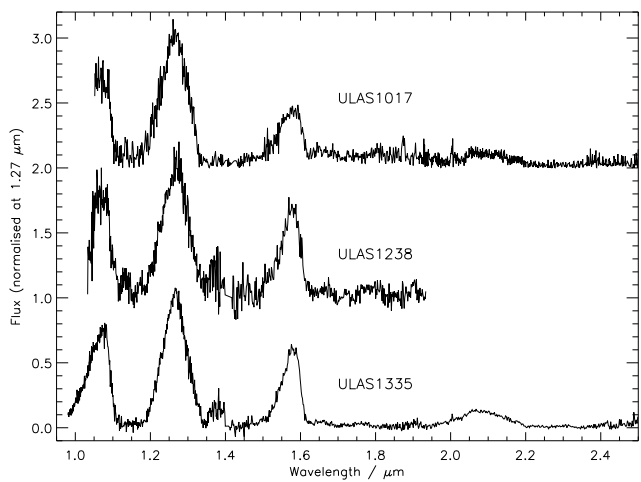}
\caption{The NIRI $JHK$ spectrum for ULAS1335 (bottom), the $JH$ spectrum
  for ULAS1238 (middle). The IRCS $J$ and NIRI $HK$ spectrum of ULAS1017
  is shown on the top row. }
\label{fig:spec1}
\end{figure}

\begin{figure}
\includegraphics[height=250pt, angle=90]{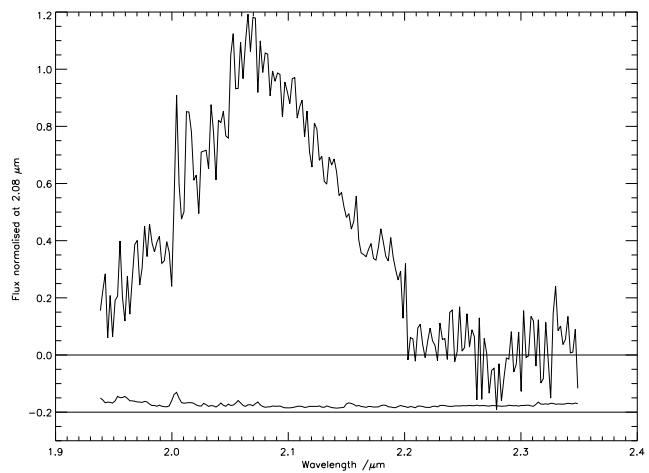}
\caption{The NIRI $K$-band spectrum for ULAS1238. The error spectrum is
  shown offset by -0.2.} 
\label{fig:1238Kspec}
\end{figure}

The spectra were then joined together
using the measured near-infrared photometry to place the spectra on an
absolute flux scale, and rebinned by a factor of three to increase the
signal-to-noise, whilst avoiding under-sampling for the spectral resolution.  
The IRCS JH spectrum of ULAS1017 was joined to the NIRI $H$ and $K$-band
spectra in a similar way, but the NIRI spectra were rebinned by a
factor of three prior to joining. 
The resultant $YJHK$ and $JHK$ spectra for ULAS1017 and ULAS1335 are shown in
Figure~\ref{fig:spec1}, where the combined spectra have been
normalised to unity at $1.27 \micron$ and each offset for clarity. 
In the case of ULAS1238, we have so far been unable to obtain a $K$-band
magnitude, so only the $JH$ spectrum is shown in
Figure~\ref{fig:spec1}, the $K$-band spectrum is shown separately in
Figure~\ref{fig:1238Kspec}, normalised to unity at $2.08 \micron$.

\subsection{Spitzer-IRAC follow-up of ULAS1335}
\label{subsec:iracobs}

We obtained IRAC four-channel (3.55, 4.49, 5.73 and 7.87\,$\mu$m)
photometry of the source on 2008 March 5.  The data were
obtained as part of the {\em Spitzer Space Telescope} Cycle 4 program
$\#40449$. All four channels have 256$\times$256-pixel detectors with a
pixel size of 1\farcs2$\times$1\farcs2, yielding a
5\farcm2$\times$5\farcm2 field of view.  Two adjacent fields were
imaged in pairs (channels 1 and 3; channels 2 and 4) using dichroic
beam splitters.  The telescope was then nodded to image a target in all
four channels.\footnote {For more information about IRAC, see
 \citet{faz04} and the IRAC Users Manual at
 http://ssc.spitzer.caltech.edu/irac/descrip.html} We used exposure
times of 30~s and a 9-position medium-sized (52 pixels) dither pattern,
for an exposure of 270s for each channel, and a total observing time of 
15.8 minutes to acquire all channels, slew, and settle.

The data were reduced using the post-basic-calibration data mosaics
generated by version 17.0.4 of the IRAC pipeline.\footnote{Information
 about the IRAC pipeline and data products can be found at
 http://ssc.spitzer.caltech.edu/irac/dh/} The mosaics were
flat-fielded and flux-calibrated using super-flats and global primary
and secondary standards observed by {\em Spitzer}. We performed
aperture photometry using an aperture with a 2-pixel (or 2\farcs4)
radius, to maximise the signal--to--noise ratio. The object was relatively 
isolated with the closest source $\sim 7$~arcsec away. 
The object appears single 
in the IRAC images, as in the case of the near-infrared images. 
To convert to total fluxes, we applied aperture
corrections as described in Chapter~5 of the IRAC Data Handbook$^3$ of
1.205, 1.221, 1.363 and 1.571
 to channels 1 through 4,
respectively. The photometry was converted from milli-Janskys to
magnitudes on the Vega system using the zero-magnitude fluxes given in
the IRAC Data Handbook (280.9, 179.7, 115.0, and 64.1 Jy for channels
1 to 4, respectively). Photometric errors were derived from the
uncertainty images that are provided with the post-basic-calibration
data.  The magnitudes and errors are given in Table~\ref{tab:iracmags}. 
Note that in addition to the errors in the table, there
are absolute calibration uncertainties of 2--3\,per cent.  There are
also systematic uncertainties introduced by pipeline dependencies of
comparable magnitude \citep{sandymix07}. We adopt the total
photometric uncertainty to be the sum in quadrature of the values
given in Table~\ref{tab:iracmags} plus 3\%.

%Note that IRAC observations are reported as a flux density at the nominal
%wavelengths given in Table~\ref{tab:iracmags}, assuming that the target has
%a flux density $f_{\nu}\propto 1/\nu$.  This assumption is not valid for
%T dwarfs and so the results given in the Table should not be used to
%derive a spectral flux at the nominal wavelength.  However, if the
%mid-infrared spectral energy distribution is known, \citet{cush06}
%show how the IRAC fluxes can be used to photometrically calibrate the
%spectrum.

 \begin{table}
\centering
\begin{tabular}{| c c |}
  \hline
Band & Magnitude (Vega) \\
$\micron$ & \\
\hline
3.55 & $15.96 \pm 0.02$ \\
4.49 & $13.91 \pm 0.01$ \\
5.73 & $14.34 \pm 0.04$ \\
7.87 & $13.37 \pm 0.07$ \\
\hline
\end{tabular} 
\caption{Mid-infrared photometry for ULAS1335. Random
  uncertainties are quoted. An additional
  3\% uncertainty should be added in quadrature to the the quoted
  random uncertainties to account for systematics.
\label{tab:iracmags}
}
 
\end{table}

\section{Spectral types}
\label{sec:spectypes}

Two methods for establishing spectral type are used. 
The first is by comparison of our spectra to the spectra of
T spectral type standards as defined by \citet{burgasser06}.
Since the latest spectral type defined by that system is T8, for which
the standard is 2MASS~J04151954-0935066 (hereafter 2MASS0415), we also
compare our spectra to the two previously identified T8+ objects,
ULAS0034 and CFBDS0059.

The $Y$ and $K$ band morphologies are more strongly dependent on
metallicity and gravity than the $J$~and $H$-bands. 
This is due to the influence of pressure dependent collisionally
induced H$_2$ absorption in the $K$~band \citep[e.g.][]{burgasser02},
and the red side of the $Y$~band flux peak \citep{settl08}. 
In addition, the blue side of the $Y$~band peak is shaped by the red wing
of the broad K{\sc I} absorption centered at 7700~\AA, which is
strongly dependent on metallicity \citep{burrows02}.
As such we base our spectral types on the comparisons of the $J$~and
$H$~band spectra of our objects. 
In this case we
normalise our $J$~and $H$-band spectra to unity at $1.27 \micron$ and
$1.58 \micron$ respectively, to facilitate comparison of their
relative shapes, rather than heights (which will be discussed separately).

The second method for establishing type is the use of the T dwarf spectral
indices, calculated as flux ratios, which are also defined by
\citet{burgasser06}. 
Aware that some spectral indices may become saturated for types later
than T8, we place our emphasis for determining type on the template
  comparison, and use this to assess the validity of the defined spectral
  indices.

\citet{delorme08} argue that ULAS0034 and CFBDS0059 may represent the
transition to Y~spectral class due to the onset of
broad NH$_3$ absorption in the blue side of the $H$-band flux peak. 
They put forward the NH$_3$--$H$ ratio as a spectral index, and
suggest that it traces the strength of ammonia absorption (see Table~\ref{tab:ratios}).
However, given the absence of calculated ammonia opacities in the
near-infrared, and possible influence of nearby water absorption band
we do not feel that this identification is certain.
Pending further confirmation of this feature, we will refer
to this index as ``NH$_3$''--$H$.
Whether the absorption feature attributed to
ammonia by \citet{delorme08} is indeed the first defining feature of
the Y~class will depend on how this feature develops towards lower
temperatures and as more objects in this regime are identified.

We calculate the $W_J$ and ``NH$_3$''--$H$
 spectral indices as put
forward by \citet{warren07} and \citet{delorme08} respectively.
The $W_J$ index traces the decreasing width of the $J$-band peak
 through the ratio of flux in a section of the blue slope of
 the peak to that in the crown. 
As such the value $W_J$ index is driven by the strength of the
 $J$-band water absorption.
These indices and the flux ratios from which they are derived are
given in Table~\ref{tab:ratios} for the three objects we have
introduced, in addition to those for the T8 standard 2MASS0415 and the
two previously discovered T8+ objects. 
The final two columns of Table~\ref{tab:ratios} give the implied
type from the template comparison and the spectral type we adopt for
each object, as discussed in more detail in the following sections.

\subsection{ULAS1017}
\label{subsec:ulas1017sptype}

\begin{figure*}
\includegraphics[height=450pt, angle=90]{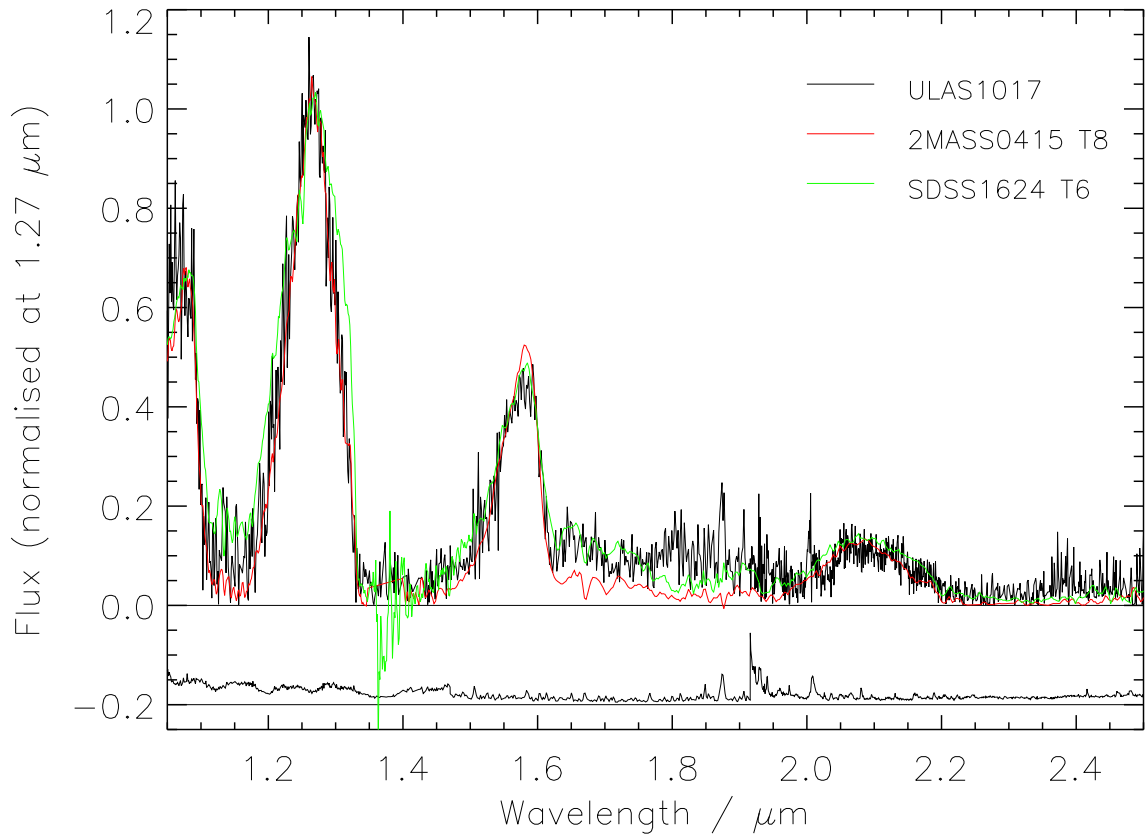}
\caption{The IRCS+NIRI $JHK$ spectrum of ULAS1017 (black line) overplotted
  with the spectra of the T8 standard 2MASS0415 and the T6 standard
  SDSS1624.
 The error spectrum, offset by -0.2 in the y-axis, is plotted as a black line.}
\label{fig:1017_comp}
\end{figure*}

The spectrum of ULAS1017 has a peculiar property. 
It appears to have a different spectral type in the $H$ and $K$-band
flux peaks as compared to the $J$-band peak. 
Figure~\ref{fig:1017_comp} shows the comparison of our IRCS+NIRI spectrum
for ULAS1017 with the T8 standard 2MASS0415 and the T6 standard
SDSS~J162414.37+002915.6 (hereafter SDSS1624).
ULAS1017 traces the form of the T8 spectrum well in the
$J$-band, but the $H$-band peak matches the T6 spectrum best, particularly
towards the the red end of this region. 
This is reflected in the calculated spectral indices for this object
(see Table~\ref{tab:ratios}).
We adopt therefore the type T8p for this object.

We do not think such morphology could arise as a result of a T8~+~T6,
or similar, unresolved binary, as flux from the brighter T6 would tend
to fill-in the deep absorption feature in the $J$-band, where we see very
little flux in this case.
Furthermore, spectral synthesis using various binary components,
following the method of \citet{burgasser08}, is not able to reproduce
the spectral ratios seen for this object. 
The problem arises because earlier type objects, which can reproduce
the $H$~and $K$~bands tend to dominate the light, and drive the ratios
in the $J$~band to earlier values also.
Although $J$-band flux reversals have been observed elsewhere
\citep[e.g.,][]{looper08}, the late spectral type of this object would
seem to preclude that interpretation. 
\citet{pinfield08} identified ULAS~J1150+0949, a T6.5p with a similarly
peculiar early $H$-band type (of T3 in this case), so such morphology
may be a generic trait for some objects.

\onecolumn
\begin{figure}
\includegraphics[height=275pt, angle=0]{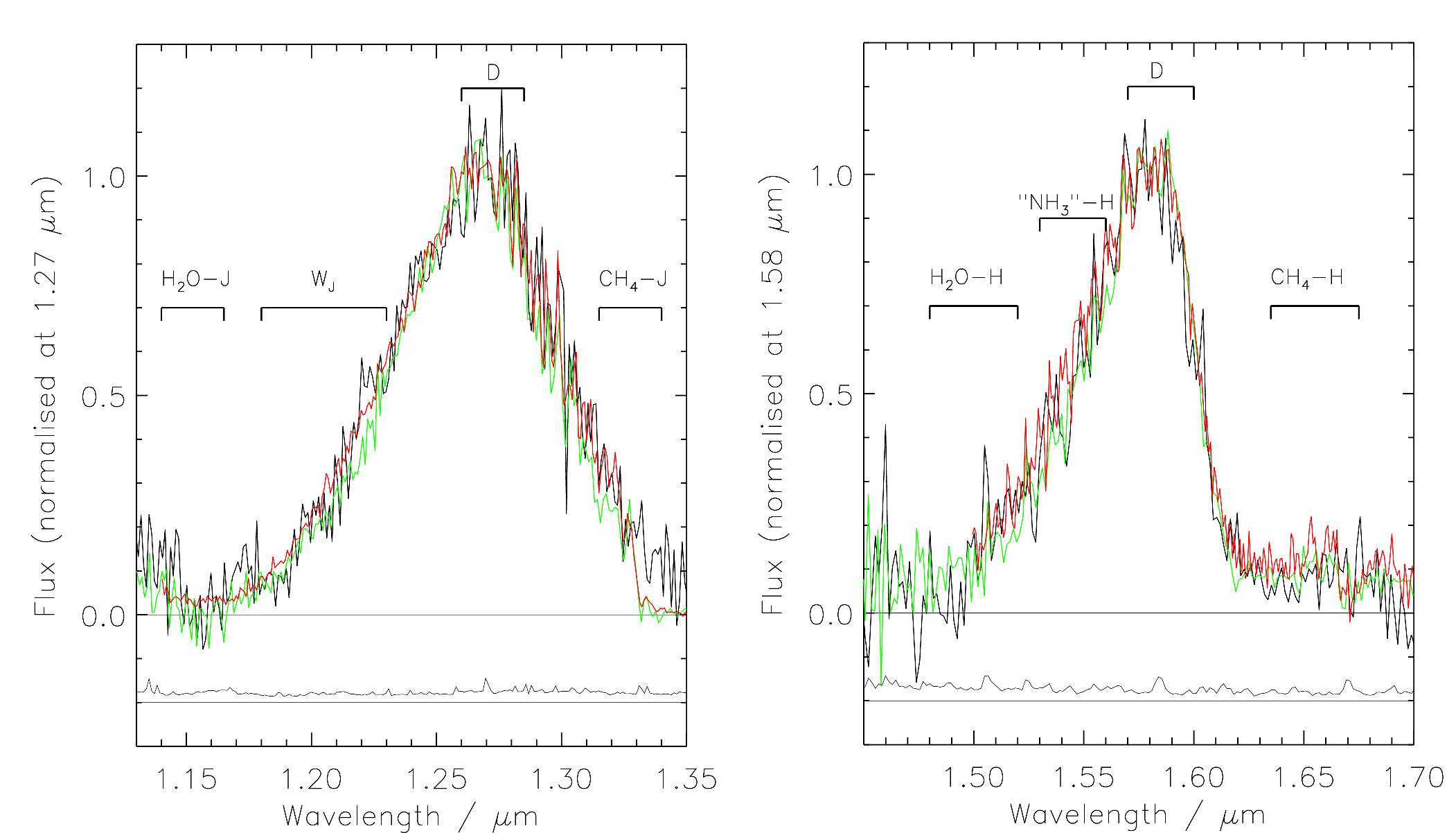}
\caption{The $J$~and $H$-band flux peaks of ULAS1238 (black line),
  ULAS0034 (T9, see text; green line) and 2MASS0415 (T8; red line). The black line that
  is plotted offset by -0.2 in the y-axis is the error spectrum for
  ULAS1238. Also indicated are the spectral regions integrated to form
  the numerator of the spectral indices discussed in the text. The
  region that forms the denominator is indicated with the letter ``D''.}
\label{fig:1238_comp}
\end{figure}

\begin{figure}
\includegraphics[height=275pt, angle=0]{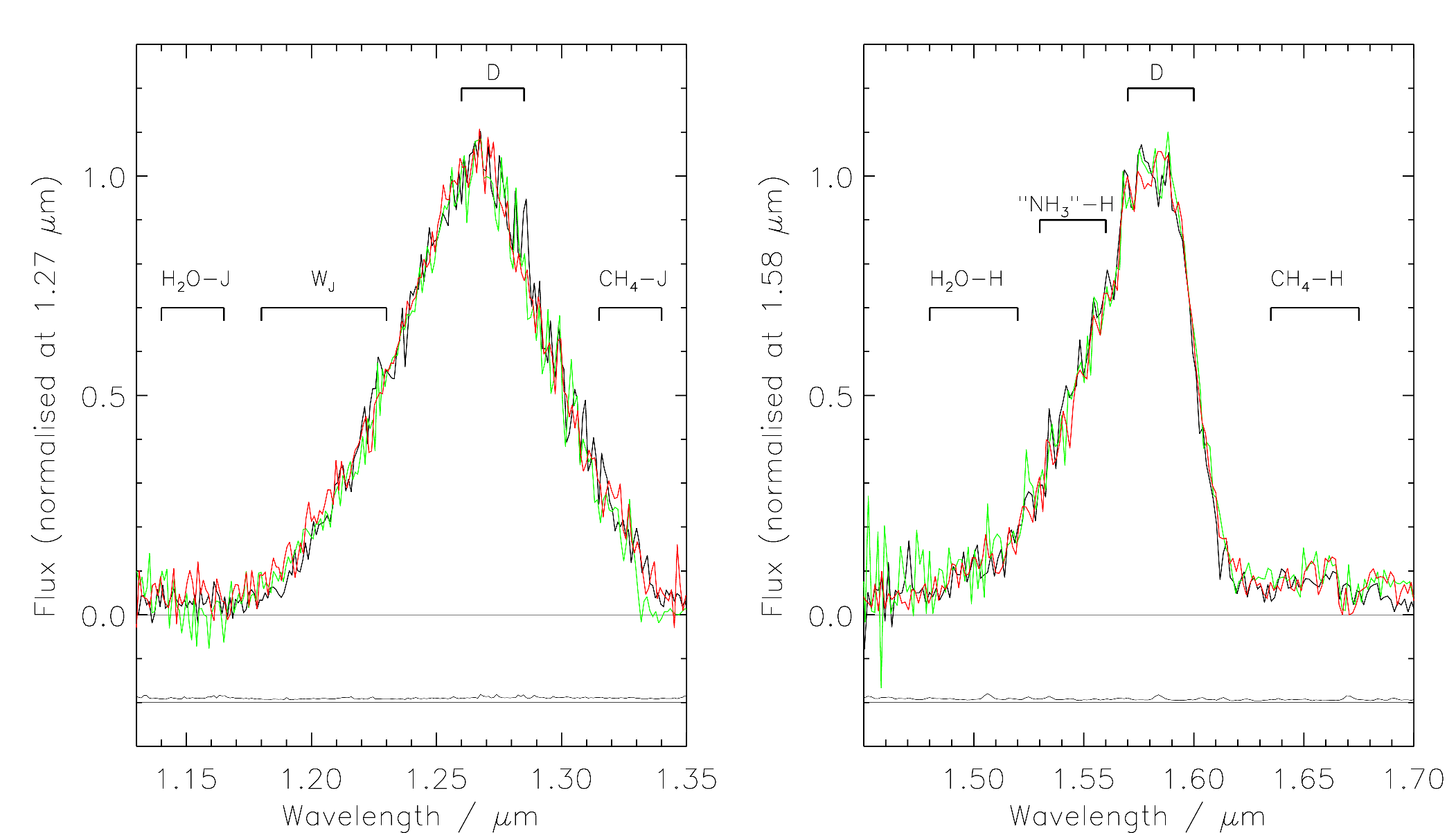}
\caption{The $J$~and $H$-band flux peaks of ULAS1335 (black line),
  ULAS0034 (T9, see text; green line) and CFBDS0059 (T9, see text; red line). The black line that
  is plotted offset by -0.2 in the y-axis is the error spectrum for
  ULAS1335. Also indicated are the spectral regions integrated to form
  the numerator of the spectral indices discussed in the text. The
  region that forms the denominator is indicated with the letter ``D''.}
\label{fig:1335_comp}
\end{figure}

\twocolumn

\begin{landscape}
\begin{table}
\vspace{6cm}
{\tiny
\begin{tabular}{c c c c c c c c c c c c c c c c}
  \hline
Index & H$_2$O - $J$ & CH$_{4}$ - $J$ & $W_J$ & H$_2$O - $H$ & CH$_4$ - $H$ & ``NH$_3$" - $H$ & CH$_4$ - $K$ & $K$ / $J$ & Template & Type \\
\hline
Flux ratio 
& $\frac{\int^{1.165}_{1.14} f(\lambda)d\lambda}{\int^{1.285}_{1.26}f(\lambda)d\lambda }$
&  $\frac{\int^{1.34}_{1.315} f(\lambda)d\lambda}{\int^{1.285}_{1.26}f(\lambda)d\lambda }$
&  $\frac{\int^{1.23}_{1.18} f(\lambda)d\lambda}{2\int^{1.285}_{1.26}f(\lambda)d\lambda }$
&  $\frac{\int^{1.52}_{1.48} f(\lambda)d\lambda}{\int^{1.60}_{1.56}f(\lambda)d\lambda }$
&  $\frac{\int^{1.675}_{1.635} f(\lambda)d\lambda}{\int^{1.60}_{1.56}f(\lambda)d\lambda }$
&  $\frac{\int^{1.56}_{1.53} f(\lambda)d\lambda}{\int^{1.60}_{1.57}f(\lambda)d\lambda }$
&  $\frac{\int^{2.255}_{2.215} f(\lambda)d\lambda}{\int^{2.12}_{2.08}f(\lambda)d\lambda }$
&  $\frac{\int^{2.10}_{2.06} f(\lambda)d\lambda}{\int^{1.29}_{1.25}f(\lambda)d\lambda }$
&
&
\\
\hline
\hline
 2MASS0415 & 0.030 (T8) & 0.168 (T8) & 0.31 & 0.173 (T8) & 0.105 (T8) & $0.625 \pm 0.003$  & 0.050 ($>$T7) & 0.134 & ... & T8 \\
 ULAS0034   & $0.012 \pm 0.006$ ($>$T8) & $0.144 \pm 0.009$ ($>$T8) & $0.262 \pm 0.006$ & $0.133 \pm 0.010$ ($>$T8) & $0.096 \pm 0.006$ (T8) & $0.516 \pm 0.008$ & $0.091 \pm 0.015$ ($>$T7) & $0.128 \pm 0.003$ & $>$T8 & T9  \\
 CFBDS0059  & $0.029 \pm 0.005$ (T8) & $0.164 \pm 0.008$ (T8) & $0.257 \pm 0.004$ & $0.119 \pm 0.008$ ($>$T8) & $0.084 \pm 0.002$ (T8) & $0.526 \pm 0.005$ & $0.128 \pm 0.037$ ($>$T7) & $0.101 \pm 0.002$ & $>$T8 & T9 \\
 ULAS1017   & $0.054 \pm 0.007$ (T8) & $0.156 \pm 0.006$ (T8) & $0.347 \pm 0.005$ & $0.274 \pm 0.016$ (T6) & $0.309 \pm 0.011$ (T6) & $0.688 \pm 0.017$ & $0.174 \pm 0.018$ (T6) & $0.119 \pm 0.002$ & T8/T6 & T8p  \\
 ULAS1238   & $0.043 \pm 0.008$ (T8) & $0.218 \pm 0.009$ (T7) & $0.284 \pm 0.006$ & $0.145 \pm 0.013$ ($>$T8) & $0.085 \pm 0.009$ (T8) & $0.601 \pm 0.013$ & $0.083 \pm 0.012$ ($>$T7) & ... & T8 & T8.5 \\
 ULAS1335   & $0.034 \pm 0.004$ (T8) & $0.179 \pm 0.004$ (T8) & $0.254 \pm 0.003$ & $0.114 \pm 0.004$ ($>$T8) & $0.077 \pm 0.003$ (T8) & $0.564 \pm 0.002$ & $0.094 \pm 0.007$ ($>$T7) & $0.134 \pm 0.001$ & $>$T8 & T9 \\
\hline
\end{tabular}
}
\caption{The near-infrared spectral index ratios as defined by
  \citet{burgasser06} for the three objects presented in this paper,
  and for the T8 standard 2MASS0415 and the previously discovered T8+
  dwarfs.  Also given is the $W_J$ index as defined by
  \citet{warren07}, and the ``NH$_3$''-$H$ index defined by
  \citet{delorme08}. The final two columns give the spectral type from template comparison in the \citet{burgasser06} scheme, and the final adopted spectral type.
\label{tab:ratios}
} 
\end{table}
\end{landscape}

\subsection{ULAS1238}
\label{subsec:ulas1238sptype}

Figure~\ref{fig:1238_comp} compares the $J$~and~$H$-band flux peaks
for ULAS1238 to those for 2MASS0415 and ULAS0034.
Initial examination of these plots, and the spectral indices given in
Table~\ref{tab:ratios}, suggest a spectral type of T8 for ULAS1238.
Closer examination of the $H$-band peak reveals some similar narrow
features in the spectra of ULAS0034 and ULAS1238.
Additionally, the width of the $J$-band peak, and the values of the
$H$-band spectral indices imply that ULAS1238 is in fact intermediate
between ULAS0034 and 2MASS0415.
If we adopt the type T9 for ULAS0034 (see
Section~\ref{subsec:ulas1335sptype}), then ULAS1238 could reasonably
be classified as T8.5.
However, the inherently small changes in near-infrared spectral
morphology in this regime make such distinctions non-critical in the
absence of additional mid-infrared spectral coverage.

\subsection{ULAS1335}
\label{subsec:ulas1335sptype}

\begin{figure}
\includegraphics[height=250pt, angle=90]{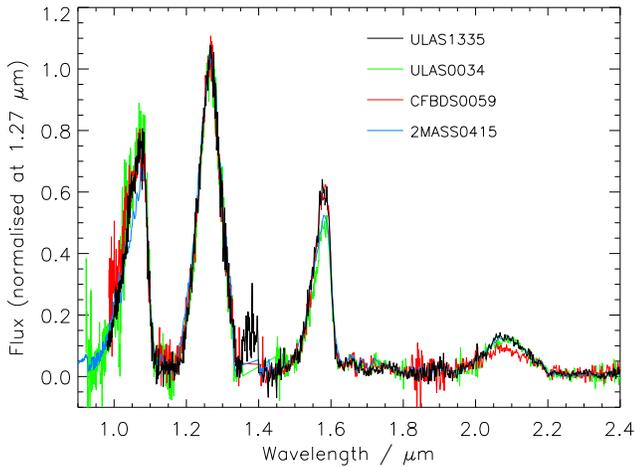}
\caption{The full $YJHK$ comparison of ULAS1335 to ULAS0034, CFBDS0059
  and 2MASS0415.}
\label{fig:1335_compfull}
\end{figure}

An initial inspection of the shapes of the $J$~and~$H$-band flux peaks
of ULAS1335 (see Figure~\ref{fig:1335_comp}) indicates that it
is of the same spectral type as ULAS0034, and CFBDS0059.
In particular it should be noted that the strength of the absorption
blueward of $1.565~\micron$ that \citet{delorme08} attribute to NH$_3$
is similar for all three objects.
Given the close similarity of these objects to the rest of the T~dwarf
sequence, it is appropriate that these objects form an extension of
that sequence, and we assign to them the subtype T9.
Figure~\ref{fig:1335_compfull} shows the full $YJHK$ comparison of
ULAS1335, CFBDS0059, ULAS0034 and 2MASS0415. 
Although the shapes of the $H$-band peaks are
similar for the three objects, their height relative to the $J$-band
peak varies significantly. 
Specifically, ULAS1335 and CFBDS0059 both exhibit much stronger $H$-band
peaks than either ULAS0034 or 2MASS0415.

\begin{figure*}
\includegraphics[height=400pt, angle=90]{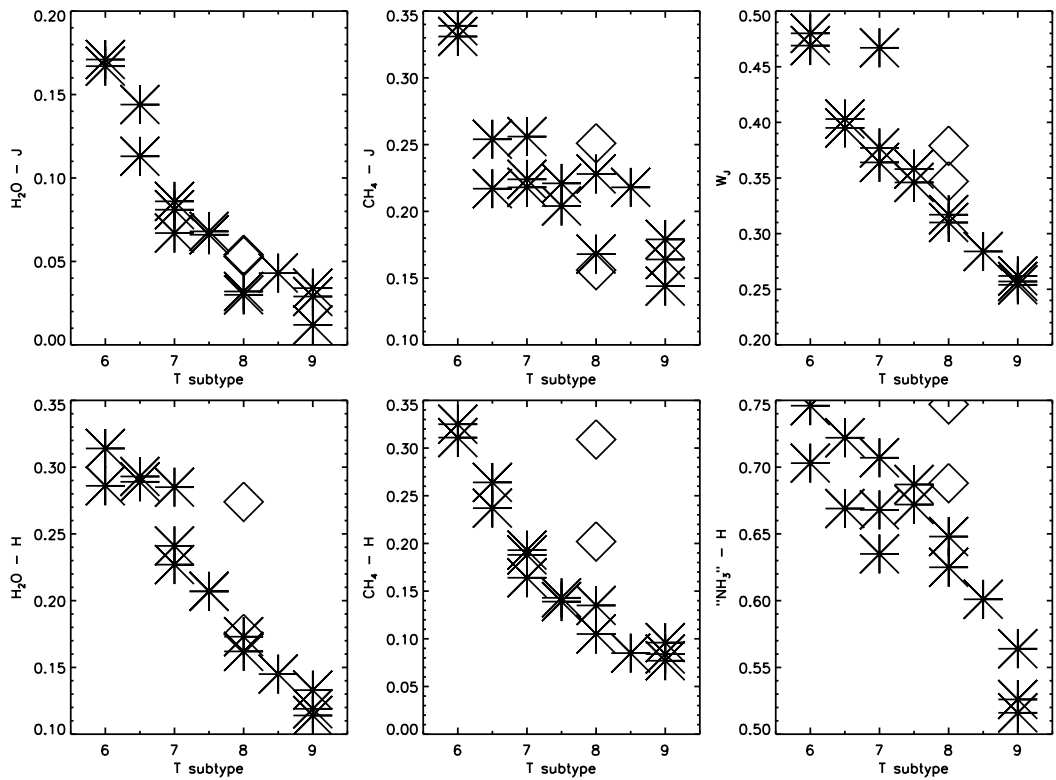}
\caption{Spectral index versus T-subtype for T6-T9 dwarfs. Asterisks
  indicate ``normal'' dwarfs, whilst the diamonds indicate two T8p
  dwarfs: ULAS1017 and 2MASS~J07290002-3954043.
Index values for these objects are drawn from
  \citet{burgasser06},\citet{warren07}, \citet{delorme08} and
  \citet{looper07}, or are calculated from the objects' spectra
  supplied by these authors. 
Uncertainties are smaller than the symbol sizes.
}
\label{fig:indices}
\end{figure*}

In Figure~\ref{fig:indices} we plot the $J$~and $H$~band spectral
indices against the spectral types derived by template comparison for
objects with types T6 to T9 (see below). 
It is apparent that the water and methane $J$-band indices previously
used for the classifying T~dwarfs are degenerate with type for the
subtypes T8 and T9.
Examination of Figure~\ref{fig:indices} indicates that the same may be
true for the CH$_4$--$H$ index, although  the H$_2$O--$H$ index is
still able to distinguish these objects. 
The new indices ``NH$_3$''--$H$ and  $W_J$ both appear able to distinguish
the T9 dwarfs from the earlier type objects, however the
``NH$_3$''--$H$ index is degenerate with type for the earlier objects.

\begin{figure}
\includegraphics[height=250pt, angle=90]{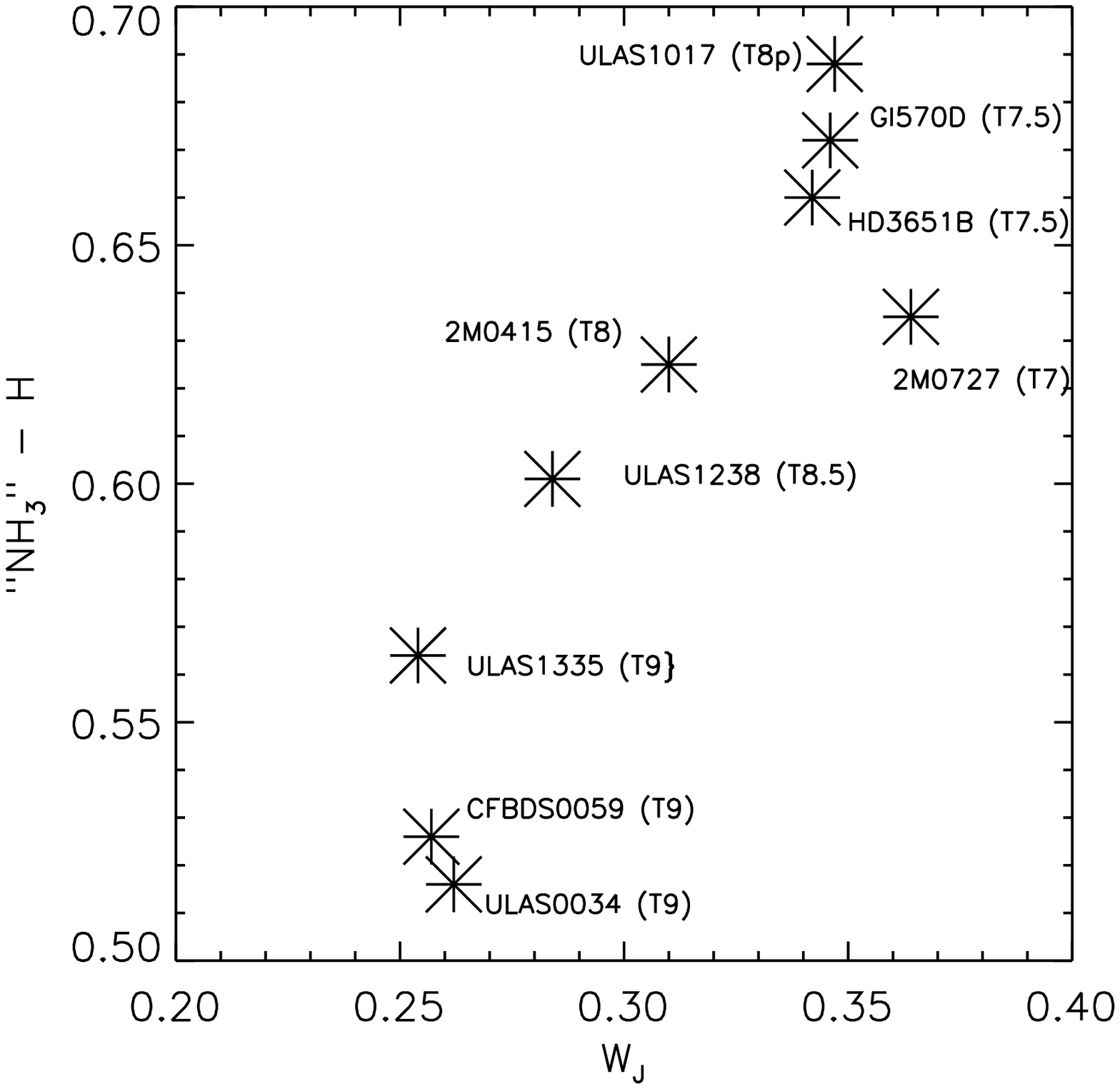}
\caption{The ``NH$_3$''--$H$ index versus $W_J$ index. Uncertainties are not
  displayed as they are smaller than the symbol size.}
\label{fig:NH3_wj}

\includegraphics[height=250pt, angle=90]{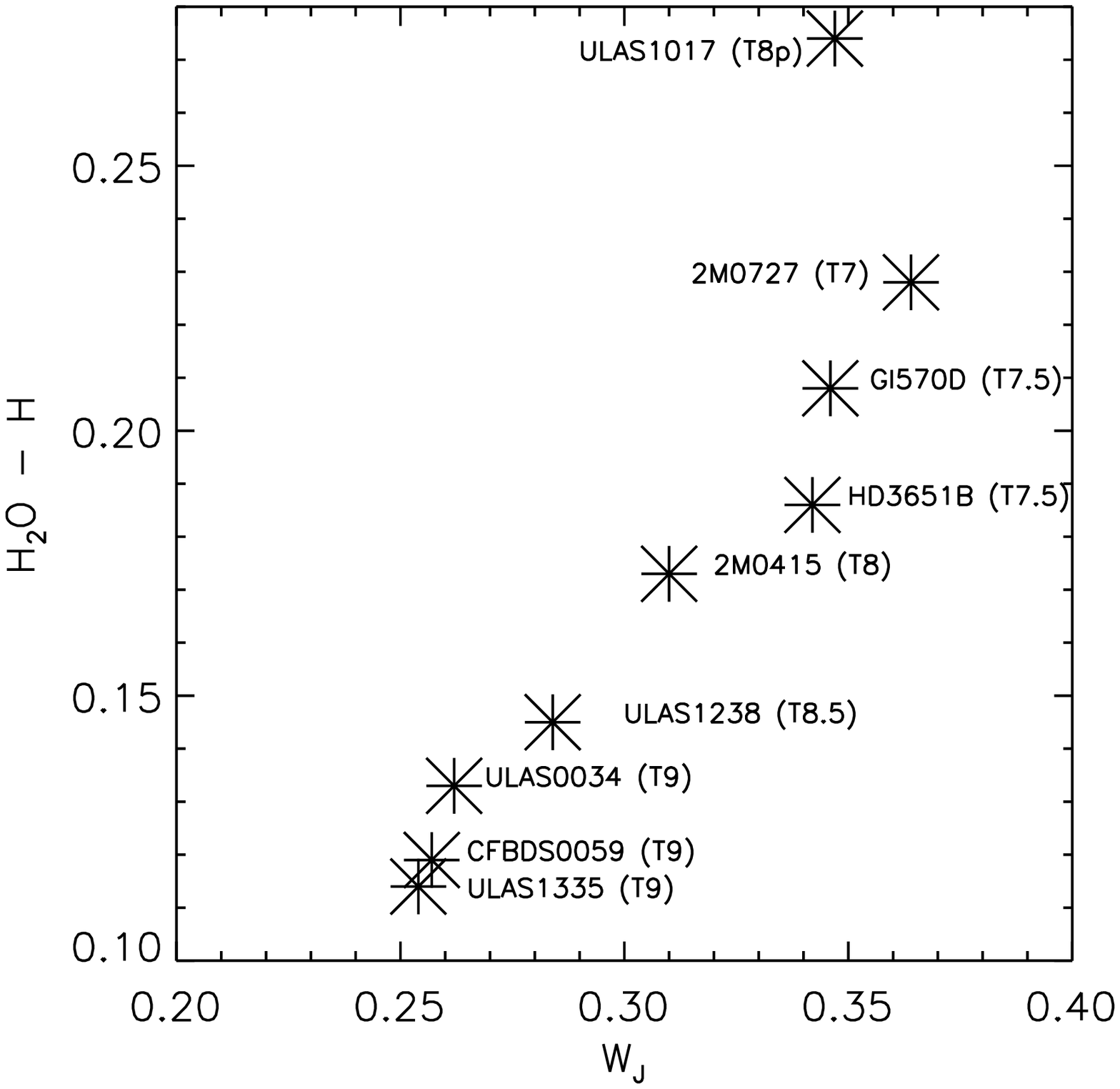}
\caption{The H$_2$O--$H$ index versus $W_J$ index. Uncertainties are not
  displayed as they are smaller than the symbol size. }
\label{fig:h2o_wj}
\end{figure}

In Figure~\ref{fig:NH3_wj} we plot the ``NH$_3$''--$H$ index against
the $W_J$ spectral index suggested by \citet{warren07}. 
It is apparent that the latest objects are well separated from
$\leq$T8 objects in this plot.
However, the values of $W_J$ and ``NH$_3$''--$H$ do not show a
consistent correlation for the latest objects.
Figure~\ref{fig:h2o_wj} shows the H$_2$O--$H$ index plotted with $W_J$. 
Again, the three latest objects are well separated from the other
objects, however in this case the values of the two indices show a
clear trend (with the exception of the T8p dwarf), both decreasing together. 
The scatter in ``NH$_3$''--$H$ values in Figure~\ref{fig:NH3_wj},
compared to the correlation shown in Figure~\ref{fig:h2o_wj} thus
suggests that although ``NH$_3$''--$H$ is useful for identifying
T8+ objects, it is not an effective typing index for very late
objects in the range we are considering, and does not conveniently
strap on to the previous T dwarf sequence.

We suggest that the appropriate spectral indices for the extension
of the T-spectral class to T9 are H$_2$O--$H$ and $W_J$. 
To provide continuity with the earlier T-indices we define the range
of $W_J$ and H$_2$O--$H$ for types T7-T9, matching our values to
the H$_2$O--$H$ values of \citet{burgasser06} for types T7 and T8, and
using the apparent correlation in Figure~\ref{fig:h2o_wj} to arrive at
appropriate $W_J$ values.
Our proposed set of spectral indices for the latter portion of
the T~spectral sequence are summarised in Table~\ref{tab:tindex}.
In this scheme, ULAS1335, as well as ULAS0034 and CFBDS0059, are all
T9 dwarfs, which are our adopted types.
Since ULAS1335 is significantly brighter than both ULAS0034 and
CFBDS0059, we also propose ULAS1335 as the template for the T9 subtype.

In Section~\ref{subsec:ulas1335_atm} we estimate a
$T_{\rm eff}$ for ULAS1335 which is marginally cooler than
the \citet{warren07} estimate for ULAS0034.
This temperature sequence is well represented by the locations
of the objects in Figure~\ref{fig:h2o_wj}, but not by the
sequence shown in Figure~\ref{fig:NH3_wj}.
If the ``NH$_3$''--$H$ index is indeed tracing a
broad ammonia absorption band, this suggests two things. 
Firstly, it appears to develop at odds with the
water absorption in the $J$-band (which the $W_J$ feature traces) and
the H$_2$O--$H$ absorption feature,
Secondly, if the relative temperatures of ULAS0034 and ULAS1335 are to
be believed, it does not correlate well
with temperature.

There is some theoretical justification for expecting near-infrared
ammonia absorption to be only weakly dependent on $T_{\rm eff}$.
Ammonia chemistry is governed by upmixing, and in practice the mixing
ratio depends most strongly on the ammonia abundance at the quench
level.
Since this is at a nearly fixed temperature, $\sim$2000~K, the remaining
variables that will affect it are pressure and composition.
Thus, we expect the ammonia mixing ratio, and by extension the
strength of ammonia absorption in the near-infrared, to be far more
sensitive to gravity and metallicity than $T_{\rm eff}$.
Given this behaviour, there is exciting potential for the
``NH$_3$''--$H$ index as an additional diagnostic tool for very cool
T~dwarfs, and its development will be followed with interest as more
more objects in the very low-temperature regime are identified.

\begin{table*}
\centering
\begin{tabular}{| c c c c c c c c|}
  \hline
NIR spectral type & H$_2$O-J & CH$_4$-J & $W_J$ & H$_2$O-$H$ & CH$_4$-$H$ & CH$_4$-K & Standard \\
\hline
T7 & 0.07--0.13 & 0.21--0.28 & 0.35--0.40 & 0.20--0.26 & 0.15--0.25 & $<$0.13 & 2MASS0727 \\
T8 & $<$0.07 & $<$0.21 & 0.28--0.35 & 0.14--0.20 & $<$0.15 & ... & 2MASS0415\\
T9 & ... & ... & $<0.28$ & $<0.14$ & ... & ... & ULAS1335 \\
\hline
\end{tabular} 
\caption{Our proposed spectral indices for the latter-most portion of
  the T~spectral sequence. Also given are the spectral type standards for each type.
\label{tab:tindex}
}
 
\end{table*}

\section{Searching for common proper motion companions}
\label{sec:propermotions}

It has been known for some time that the degree of multiplicity 
amongst very young stars is greater than that of the more evolved 
field star populations \citep{duq91,leinert93}, 
and thus that the majority of binary systems form together in their 
nascent clouds. Binary components can therefore generally be assumed 
to share the same age and composition, and wide common proper motion 
companions to our new objects could thus provide valuable constraints 
if they are to be found. In order to address this possibility and 
more generally assess the kinematics of our objects, we calculated 
proper motions for our targets using the highest signal-to-noise 
multi-epoch observations available. In each case this was the WFCAM 
and UFTI $J$-band images.

The IRAF routines geomap and geoxytran were used to transform
between the available multi-epoch images, using an average
of 12 reference stars. This allowed any
motion of the T dwarfs to be accurately measured. Proper motion 
uncertainties were estimated from centroiding accuracies combined 
with the residuals associated with our derived transformations. 
The calculated proper motions are given in Table~\ref{tab:motions}.
We also estimate $V_{tan}$ based on our distance estimates derived in
Section~\ref{sec:fundies}. These estimates lie within the range of
$V_{tan}$ observed by \citet{vrba04} for L~and T~dwarfs.

We independently searched the Supercosmos Sky Survey database, and 
the Simbad database (accessing the Hipparcos and Tycho catalogues 
amongst others), for common proper motion companions (within 1$\sigma$) 
of the three new T dwarfs, thus searching for wide binary companions 
over a wide range of mass. The search was made within a projected 
distance of 20,000 AU from each target, calculated assuming the 
minimal distance within our derived constraints (see
Section~\ref{sec:fundies})
to ensure a spatially complete search.
In the case of ULAS1238 we used a lower distance estimate of 20~pc based on
comparison with ULAS0034.
 SDSS and UKIDSS photometry 
were then retrieved for any common proper motion objects, and from 
these a spectral type range was estimated, and an approximate distance 
modulus inferred. Sources with distance moduli that were inconsistent 
with the inferred distances to our newly discovered dwarfs (see 
Section~\ref{sec:fundies}) could then be rejected as potential
companions. Our search produced no potential companions for any of our targets. 

 \begin{table*}
\centering
\begin{tabular}{| c c c c c c c c|}
  \hline
Object & \multicolumn{2}{|c|}{1st Epoch Coordinates} & Epoch 1 & Epoch 2 & $\mu_{\alpha cos \delta}$ & $\mu_\delta$ & $V_{tan}$\\
 & $\alpha$ & $\delta$ & & & mas yr$^{-1}$ & mas yr$^-1$ & kms$^{-1}$ \\
\hline
ULAS1017 & 10 17 21.40 & +01 18 17.9 & 2007-03-17 & 2008-01-17 & $110 \pm 80$ & $40 \pm 80$ & 4-58 \\
ULAS1238 & 12 38 28.51 & +09 53 51.3 & 2007-02-13 & 2008-01-25 & $418 \pm 40$ & $21 \pm 40$ & ... \\
ULAS1335 & 13 35 53.45 & +11 30 05.2 & 2007-04-21 & 2008-01-16 & $160 \pm 50$ & $-290 \pm 50$ & 10-21 \\

\hline
\end{tabular} 
\caption{Proper motion estimates for the three new T
  dwarfs. Uncertainties reflect the residuals in the transformations
  between the two images in each case. 
Also given are estimates for $V_{tan}$ for ULAS1335 and ULAS1017 (see
  Section~\ref{sec:fundies} for distances). 
\label{tab:motions}
}
 
\end{table*}

\section{Atmospheric parameters}
\label{sec:atm}

In the following two sections we estimate the atmospheric
parameters of ULAS1017 and ULAS1335 through model comparisons, and, in the case
of ULAS1335, consideration of the near-infrared to mid-infrared
spectral energy distribution (SED). The model spectra we employ 
are a recent realisation of the BT-Settl models, for which detailed
descriptions of the input physics are given in \citet{warren07} and \citet{settl08}. 
There 
are uncertainties in these models that limit their use 
significantly for simplistic comparisons, but by limiting our analysis 
to relative comparisons over key spectral ranges, and some extrapolation 
of theoretical trends, we are able to derive a set of ``best-guess'' 
properties based on the existing theoretical and observational data.

We do not perform a detailed analysis for ULAS1238 due to the
absence of $K$-band photometry that precludes flux calibration for 
this part of the spectrum. Instead we speculate that its properties 
are likely intermediate between 2MASS0415 and ULAS0034 since its 
spectral type is also intermediate. However, we do not rule out the
possibility of a considerably lower $T_{\rm eff}$ (see Section~\ref{sec:jup}).

\subsection{ULAS1017}
\label{subsec:ulas1017_atm}

To estimate the atmospheric parameters for ULAS1017 we repeat the
($W_J$,$K/J$) analysis described by \citet{warren07} for
ULAS0034, and applied to CFBDS0059 by \citet{delorme08}.
To this end we have calculated $W_J$ and $K/J$ indices for a grid of
solar metallicity BT-Settl models with  $4.20<$~log~{\it g}~$<5.50$, and $600K
< T_{\rm eff} < 900K$.
We have anchored the resulting ($W_J$,$K/J$) grid such that $T_{\rm
  eff}=750K$ and log~{\it g}~=~5.0 for the empirical ($W_J$,$K/J$)
coordinates of 2MASS0415, consistent with the detailed analysis of
\citet{saumon07}. 
Placing ULAS1017 on this grid implies a $T_{\rm eff} = 800$K and
log~{\it g}=5.5.

\begin{figure}
\includegraphics[height=250pt, angle=90]{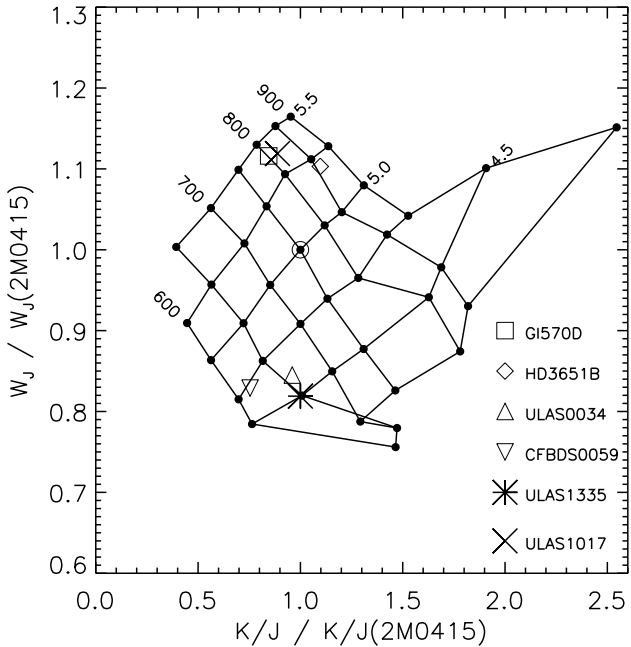}
\caption{$W_J$ versus $K/J$ indices for a grid of solar metallicity
  BT-Settl model spectra. The grid has been normalised by matching the
  $T_{\rm eff} = 750K$, log~{\it g}~=~5.0 model indices to the empirical
  index values for 2MASS0415, which lies at (1.0, 1.0). }
\label{fig:kjwj_sol}
\end{figure}

\begin{figure}
\includegraphics[height=350pt, angle=0]{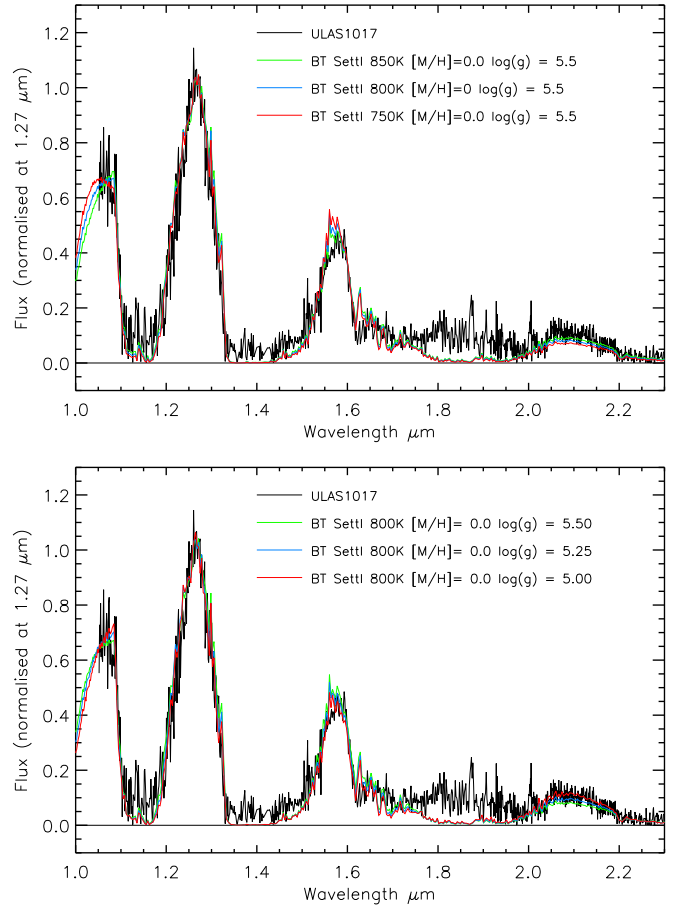}
\caption{A comparison of model spectra for the parameters bracketing the
  location of ULAS1017 in the solar metallicity ($W_J$,$K/J$) plane.}
\label{fig:1017models}
\end{figure}

In Figure~\ref{fig:1017models} we compare the spectrum of ULAS1017 to
  model spectra that bracket its location in the ($W_J$,$K/J$) plane.
It appears that the model spectra generally fit the spectrum of
  ULAS1017 well.
We note, however, that it is the model for $T_{\rm eff} = 800$K and
  log~{\it g}~=~5.0 that provides the best fitting spectrum, in contrast to
  the ($W_J$,$K/J$) analysis that suggests log~{\it g}~=~5.5.

The location of ULAS1017 in Figure~\ref{fig:kjwj_sol} is very close to
  that of Gl~570D.
It is worth noting that using detailed analysis with consideration of
  non-equilibrium effects,\citet{saumon06} derive parameters for Gl~570D
  of $T_{\rm eff} = 800-820$K and log~{\it g}~=~5.09-5.23. 
This is very similar to our best value from spectral
  comparison for ULAS1017, but similarly at odds with the parameters
  from simple ($W_J$,$K/J$) analysis.

Note also that this form of analysis can only provide
approximate estimates of the atmospheric parameters, and it must be
remembered that varying estimates of surface gravity will be
degenerate with respect to varying metallicity.
The lack of $Y$-band coverage in our spectrum of ULAS1017 prevents us
from making any estimate of its metallicity relative to other objects
based on the form of its $Y$-band peak.
Furthermore, the form of the $H$-band flux peak, resembling that of a T6
dwarf adds an additional element of uncertainty to our ``best-guess''
properties for this object.

\subsection{ULAS1335}
\label{subsec:ulas1335_atm}

\begin{figure}
\includegraphics[height=350pt, angle=0]{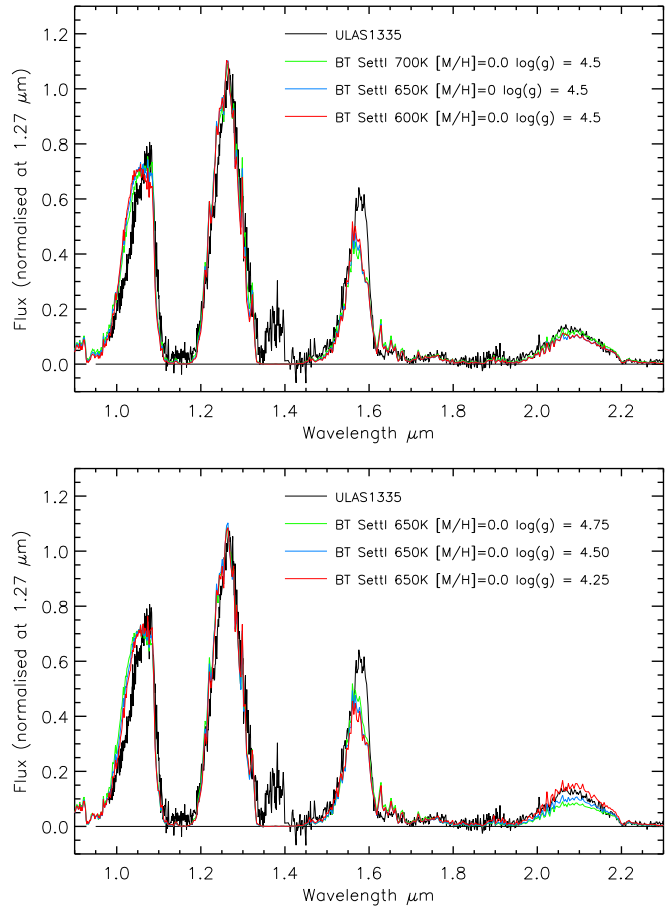}
\caption{A comparison of model spectra for parameters bracketing the
  location of ULAS1335 in the solar metallicity ($W_J$,$K/J$) plane.}
\label{fig:models1}
\end{figure}

We will begin by placing ULAS1335 on the ($W_J$,$K/J$) grid (see Figure~\ref{fig:kjwj_sol}), 
which implies $T_{\rm eff} \approx 650$K and log~{\it g} $\approx 4.5$. To 
assess the credence of these theoretical predictions, Figure~\ref{fig:models1} 
shows the comparison of ULAS1335 with the BT-Settl synthetic spectra for the 
atmospheric parameters that bracket its location in the ($W_J$,$K/J$) plane 
for solar metallicity. It is clear that these synthetic spectra are not a 
good fit to the data. The most obvious discrepancy is the height of
the $H$-band flux peak.  
It is widely acknowledged that the shape of the $H$-band peak is 
poorly reproduced by model spectra, a fact that is attributed to incomplete 
methane line lists. However, in the cases of ULAS0034, CFBDS0059 and
2MASS0415, the  observed height of the $H$-band flux peak is well
represented by the model  spectra for the atmospheric parameters
estimated from their location in the  ($W_J$,$K/J$) plot
\citep[see][figures 6 and 7 respectively]{warren07,delorme08}
It is thus clear that despite many similarities 
between ULAS1335 and the other two T9 dwarfs, its $H$-band peak is, by
comparison, significantly brighter. Some of the properties of ULAS1335
are thus presumably somewhat different to the other two T9 dwarfs.

We can make a first assessment of possible differences in metallicity by 
considering the shape of the $Y$-band peak. All available theoretical models 
indicate that the shape of this flux peak will be sensitive to [M/H], although 
there is some ambiguity as to exactly how compositional changes will
affect its  shape. A comparison of the $Y$-band peaks of the three T9
dwarfs in Figure~\ref{fig:1335_compfull}  reveals them to have an
extremely similar $Y$-band spectral morphology, and it  thus seems
unlikely that these objects have significantly different metallicity.

Figure~\ref{fig:models1} shows what changes we might expect for
varying $\log{\it g}$ and $T_{\rm eff}$. 
In the bottom plot it can be seen that although a higher 
gravity can increase the brightness of the $H$-band peak relative to the $J-$ 
and $Y$-band peaks, it has the opposite effect on the $K$-band flux peak. 
These trends thus indicate that it is not possible to theoretically account 
for the relatively bright $H$-band peak of ULAS1335 compared to the 
other T9 dwarfs, simply by varying $\log{\it g}$ alone.

We can obtain a better match between theoretical and observed spectra if we 
adopt a lower $T_{\rm eff}$ for ULAS1335. The top plot in
Figure~\ref{fig:models1}  demonstrates the trend in the BT-Settl model
spectra, and shows that the relative  
brightness of the $H$-band peak increases compared to the $Y$- and
$J$-band peaks as $T_{\rm eff}$ decreases below 700K. However, note that
although there is some change to the relative strength of the $K$-band
peak, it is not as drastic as when one varies $\log{\it g}$. 
The overall near infrared morphology of ULAS1335 
is thus best explained (through current theory) if this object has a somewhat 
lower $T_{\rm eff}$ than ULAS0034.

\begin{figure}
\includegraphics[height=350pt, angle=0]{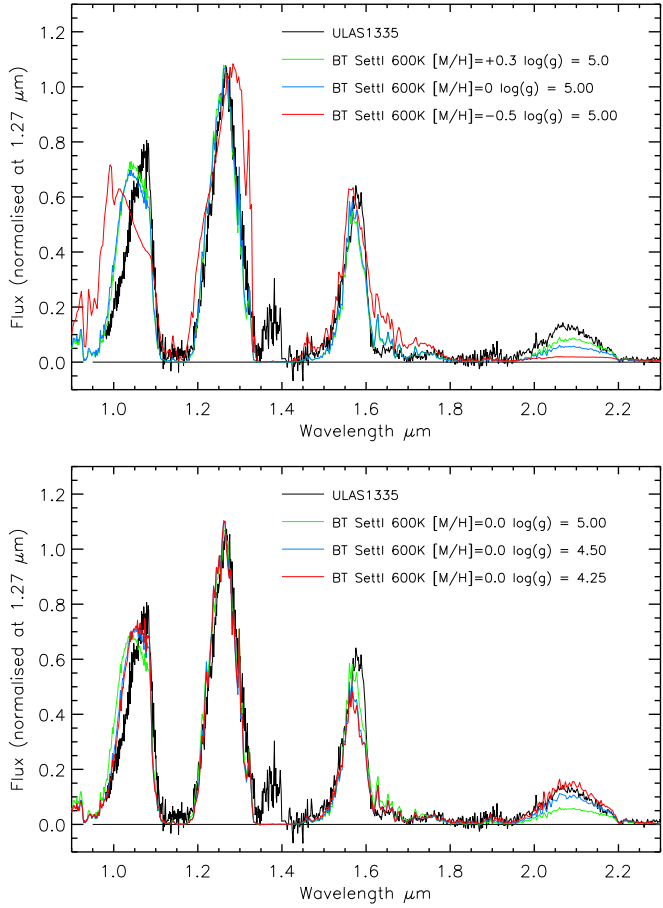}
\caption{A comparison of model spectra with ULAS1335 for differing
  gravity and metallicity combinations at $T_{\rm eff} = 600$K.}
\label{fig:models2}
\end{figure}

In Figure~\ref{fig:models2} we compare the spectra of ULAS1335 to a range of 
model spectra with different metallicities and gravities for our
lowest available 
temperature models, $T_{\rm eff} = 600$K. In the top plot, the combination 
log~{\it g}~=~5.0, $T_{\rm eff} = 600$K model spectra all estimate the height 
of the $H$-band peak well. The sub-solar metallicity model is a poor match 
everywhere else however. The high-metallicity model gives a reasonable 
approximation to the observed spectrum, but as expected it under-estimates 
the $K$-band flux somewhat, and has a significantly wider and
blue-ward shifted  $Y$-band peak.

In the bottom plot, the $\log{\it g}=$4.25-4.50 model spectra provide
a reasonable  fit to the whole near infrared spectra. However, whilst
reproducing the $Y$-, $J$-,  and $K$-band peaks, these spectra still
somewhat under-estimate the brightness of the $H$-band peak. 
Although the low $T_{\rm eff}$ of this model has gone some way to 
addressing the relative brightness of this peak, the model trends
suggest that  we should obtain a better fit by going to even lower
$T_{\rm eff}$.

\begin{figure}
\includegraphics[height=250pt, angle=0]{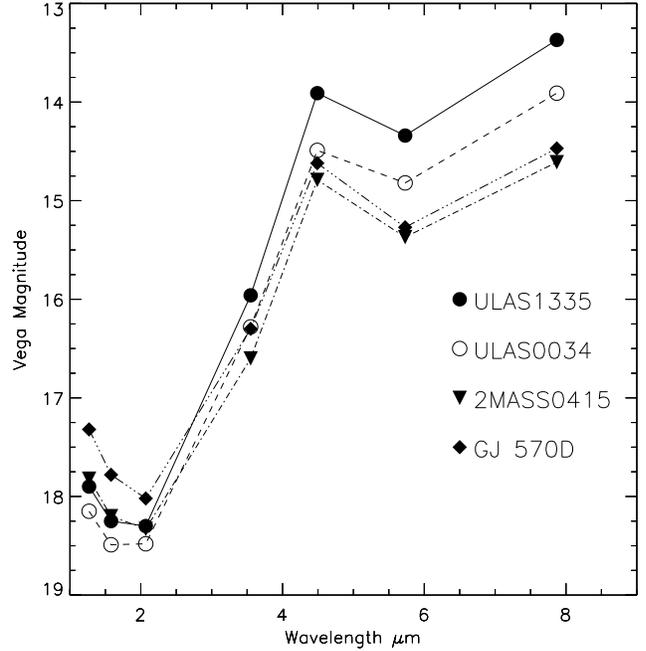}
\caption{The 1-8$\micron$ SEDs for ULAS1335,ULAS0034 \citep{warren07},
  2MASS0415 and Gl~570D. The photometry for the latter two objects, taken from
  \citet{sandy02} and \citet{patten06}, has been offset by
  +2.5 magnitudes to allow comparison with the T8+ dwarfs. 
Uncertainties
are of comparable size to the symbols.}
\label{fig:NIRMIRsed}
\end{figure}

\begin{figure}
\includegraphics[height=250pt, angle=90]{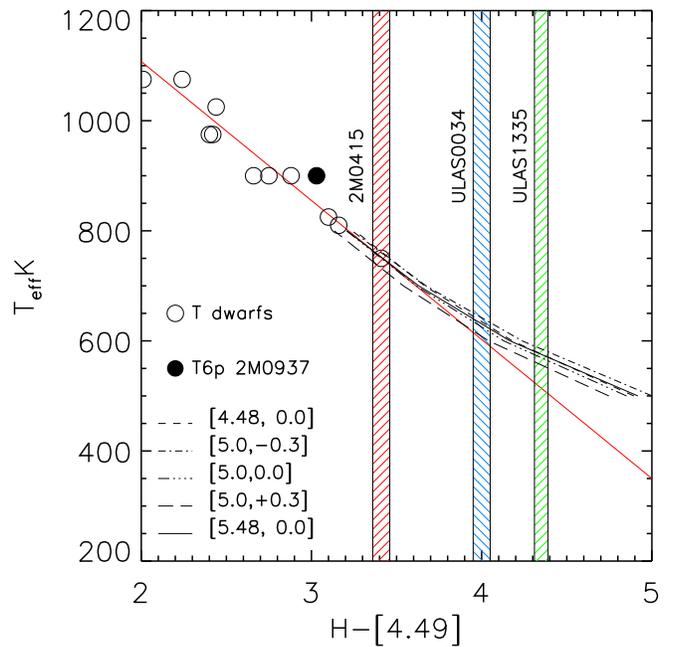}
\caption{The $T_{\rm eff}$ - $H$--[4.49] relation for T dwarfs as defined
by \citet{warren07}. The shaded regions indicate the colours ($\pm
1\sigma$) for the coolest T dwarfs with IRAC magnitudes.The straight
red line indicates the empirical fit the colours of warmer
T~dwarfs. Also plotted, as black lines are theoretical tracks from
\citet{marley02} and \citet{ms08}. Values for $\log {\it g}$ and
metallicity are indicated in brackets.}
\label{fig:H449}
\end{figure}

Investigation of the near-infrared to mid-infrared SED for ULAS1335
reveals that is extremely red.
Figure~\ref{fig:NIRMIRsed} 
shows the 1-8$\micron$ SEDs for ULAS1335, ULAS0034, 2MASS0415 and
Gl~570D. 
Whilst the form of 
the SEDs are similar, ULAS1335 is significantly redder than ULAS0034 over 
the 2 - 4.49 $\micron$ range. 
ULAS1335 has $H-[4.49] = 4.34 \pm 0.04$, the reddest value yet
observed for a T~dwarf. 
\citet{warren07} identified a correlation 
between the $H$--[4.49] colour and $T_{\rm eff}$ for T dwarfs with well 
determined $T_{\rm eff}$ \citep{golim04}. In fact the correlation 
for T dwarfs with $T_{\rm eff}=800-1250$K is particularly tight, with an 
RMS about the best-fitting linear relation of $\pm$28K. 
In Figure~\ref{fig:H449} 
we show this linear fit, and indicate the $H$--[4.49] colours of 2MASS0415, 
ULAS0034 and ULAS1335. 
Theoretical tracks are also over-plotted for a recent
realisation of the \citet{marley02} model atmospheres for the range
500--800K \citep{ms08}.
Tracks are shown for $4.48 \leq \log{g} \leq 5.48$ and $-0.3 \leq
[M/H] \leq +0.3$,
bracketing the expected properties of ULAS1335.  
We have anchored the models for $\log {\it g} = 5.0$ and solar
metallicity to the colours of the coolest T~dwarfs with well
determined temperatures. 
It can be seen that the theoretical models are reasonably supportive of an 
extrapolation of the relation between $H$-[4.49] and $T_{\rm eff}$ 
for such $\log{g}$ and [M/H] ranges, and this extrapolation suggests that 
ULAS1335 has $T_{\rm eff}\sim$550--600K.

In the case of ULAS0034, the temperature inferred by ($W_J$,$K/J$)
 analysis was reasonably
consistent with that inferred from an extrapolation of the
$T_{\rm eff}$--$H$--[4.49] relation. 
However, in the
case of ULAS1335 we have seen that the two estimates differ by
 $\sim 100{\rm K}$.
The poor agreement between the model spectra for 650~K and the
observed spectrum of ULAS1335 cast doubt on the temperature
derived from the ($W_J$,$K/J$) method.
It is likely that this disagreement is principally due to the fact that T9
dwarfs are in the temperature range near the limit of where the
near-infrared model spectra are valid. 
Since no model spectra cooler than 600~K are plotted, a degeneracy
in ($W_J$,$K/J$) for these temperatures cannot be ruled out.
Indeed, Figure~\ref{fig:kjwj_sol} indicates that there is a predicted
 distortion in the grid (which will result in some degeneracy) for the lowest
gravity and temperature combinations currently considered.

The strong, relatively tight $T_{\rm eff}$ correlation offered by observed
T~dwarf $H$--[4.49] colours combined with the theoretical expectations
for a reasonably steady shift of flux from the near-infrared into the
mid-infrared with decreasing $T_{\rm eff}$, lends weight to this colour as
a better temperature indicator for such very late objects, and we adopt
the temperature implied by this colour rather than that implied by
the ($W_J$,$K/J$) method. 
We also suggest that the ($W_J$,$K/J$) method be used with caution in the
$T_{\rm eff}$ regime near the limit of current atmospheric models.

To summarise, our comparison to model near infrared spectra suggested that 
ULAS1335 has $T_{\rm eff} < 600$K. This is clearly supported by the
the near-to-mid infrared SED, which suggests $T_{\rm eff} \sim 550-600{\rm K}$ from a theoretically informed extrapolation.
Our ``best-guess'' for the temperature of ULAS1335 is thus $T_{\rm
  eff}\sim$550--600K.

This temperature is stated along with an important caveat. 
We note that preliminary investigations of recent
realisations of the \citet{marley02} model spectra \citep{ms08} do not
show the same trends with varying parameters in the near-infrared as
the BT-Settl models. 
Although we defer a detailed comparison of these model spectra with
the BT-Settl models to a future work, we highlight those differences of
particular relevance here.
Most notable is the absence of a trend of increasing height of the
$H$~band peak with decreasing $T_{\rm eff}$. 
In addition the height of the $H$~band peak appears more strongly
dependent on metallicity than in the BT-Settl models, increasing with
decreasing metallicity.

Finally, considering the absence of mid-infrared photometry for CFBDS0059, and
the reliance on the ($W_J$,$K/J$) method for its temperature estimate,
we also note that no strong conclusion regarding its temperature
relative to ULAS1335 and ULAS0034 should be drawn. We cannot rule out
the possibility that CFBDS0059 is cooler still.

\section{Comparison with the Jovian spectrum}
\label{sec:jup}

\begin{figure*}
\includegraphics[height=400pt, angle=0]{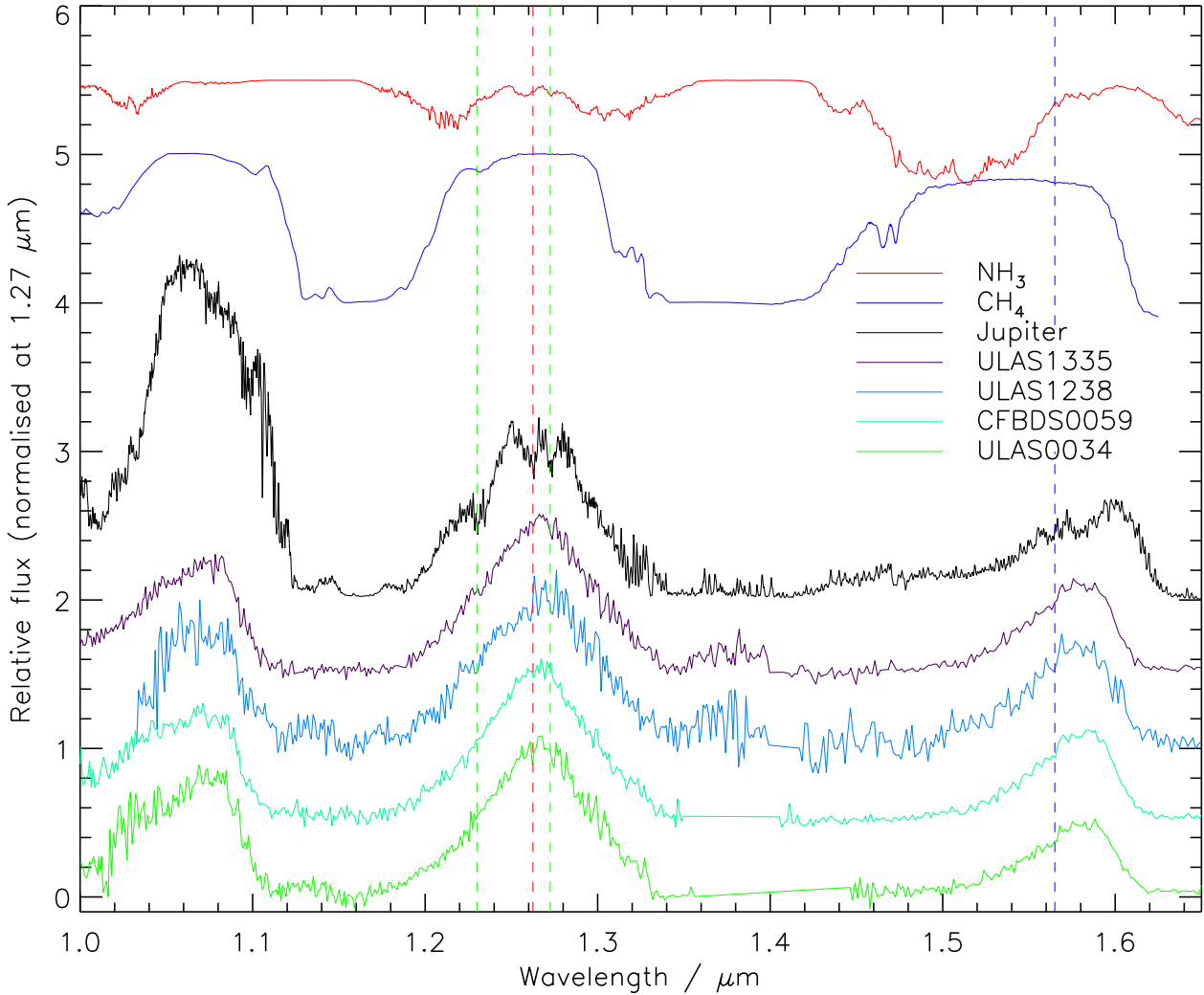}
\caption{Spectra for ULAS0034, CFBDS0059, ULAS1238 and ULAS1335
  compared with the near-infrared spectrum of Jupiter(J. T. Rayner,
  M. C. Cushing, \& W. D. Vacca 2008, in preparation) and laboratory
  transmission spectra of methane \citep{cb69} and ammonia \citep{nh3}.  
The vertical green dashed lines highlight narrow features that appear to
  be common to ULAS1238, ULAS1335 and Jupiter. The vertical red
  dashed line highlights an interesting feature in the Jovian spectrum
  that is absent from the T dwarf spectra (see text). 
The vertical blue dotted line indicates the location of the red edge
  of the broad absorption attributed to ammonia by \citet{delorme08}.  }
\label{fig:jupplot}
\end{figure*}

We have also examined our T dwarf spectra in more detail, to try and
identify new features that might be useful in this new $T_{\rm eff}$
regime.
We follow \citet{sandy07} and use the spectrum of Jupiter along with
transmission spectra of ammonia and methane as a basis for comparison,
which we plot in Figure~\ref{fig:jupplot} alongside ULAS0034,
CFBDS0059,ULAS1238 and  ULAS1335.
Although the metallicity of the Jovian atmosphere is much higher, and
the gravity much lower, than that expected for our T~dwarfs, this
comparison with a much cooler atmosphere is still interesting.

We draw attention to the narrow feature near 1.23$\micron$ (left-most
vertical green line in Figure~\ref{fig:jupplot}) which
appears to be shared by Jupiter, ULAS1238 and ULAS1335. 
This should not be confused with the K~{\sc I} doublet at $\sim 1.24 -
1.25 \micron$.
We attribute this feature to methane, based on comparison with the
laboratory transmission spectrum of \citet{cb69}.
We also highlight the narrow feature at $\sim 1.272 \micron$ (right-most
vertical green line), which also appears to be common between ULAS1238,
ULAS1335 and Jupiter. 
This may be due to ammonia, based on comparison to the
transmission spectrum of \citet{nh3}.
We note, however that the sharp feature blueward of this (indicated by a
vertical red line); also likely due to ammonia is
apparently absent from the spectra of the T~dwarfs.

\section{Physical parameters}
\label{sec:fundies}
Since metallicity is thought to only weakly impact the derivation of
fundamental parameters from $T_{\rm eff}$ and log~{\it g}
\citep[e.g.,][]{saumon07}, we use the solar metallicity relations
summarised by \citet{burrows2001} to estimate the mass, age and radius
for ULAS1017 and ULAS1335 from our $T_{\rm eff}$ and log~{\it g} estimates. 

For ULAS1017, the extreme cases from the model comparison, $T_{\rm
  eff} = 750 - 850$K and log~{\it g}~=~5.0--5.5 correspond to an age of
  1.6--15~Gyr and a mass of 33--70~M$_J$.
By estimating its radius from the same set of relations, and
  normalising the corresponding atmospheric models to the $J$-band flux
  peak, we can estimate the distance to ULAS1017 to be in the range
  31-54pc.

For ULAS1335, we use our best estimate of $T_{\rm eff}~=~550-600$K
and, since we suspect low-gravity for this object, we adopt a conservative
range of $\log~{\it g}~=~4.5-5.0$ for its surface gravity.
These atmospheric parameters only weakly constrain the age to a range
of 0.6--5.3~Gyr and the mass to 15--31~M$_J$.

In the absence of model spectra for $T_{\rm eff}~<~600$K, we have
  estimated the bolometric flux for ULAS1335 using a pseudo-synthetic
  spectrum.
This was constructed by bracketing our flux-calibrated YJHK
  spectrum with the model spectrum for $T_{\rm eff} = 600$K and
  log~{\it g}~=~4.5, scaling the synthetic parts with our $z$~band
  photometry in the optical and our IRAC [4.49] photometry in the
  mid-infrared.  
We then integrated this this pseudo-synthetic spectrum over all
  wavelengths to estimate the bolometric flux from ULAS1335.
The uncertainty in this estimate is dominated by uncertainty in the
  photometry used to calibrate the spectrum.
These uncertainties range from $10 \%$ for the optical photometry
  to $<5\%$ for our near-infrared photometry.
These regions contribute $\sim 0.5\%$ and $\sim 25\%$ to the
  bolometric flux estimate respectively. 
The mid-infrared, which contributes $\sim 75\%$ of the bolometric flux, has an
  uncertainty of $4\%$ in its [4.49] magnitude, which we use to scale
  the synthesised spectrum.
Allowing for additional uncertainty in the model spectrum match to the true
  mid-infrared spectrum, we estimate a total uncertainty in our
  bolometric flux to be $\sim 15\%$, and the bolometric flux to be
  $F_{bol} = 3.35 \pm 0.5 \times 10^{-16}$~Wm$^{-2}$.

The uncertainty in our distance estimate for ULAS1335 is dominated by
the uncertainty in $T_{\rm eff}$ (and thus luminosity).
For $T_{\rm eff}~=~550-600$K  and log~{\it g}~=~4.5--5.0  we estimate a
luminosity of $-6.18~\leq~\log~(L_* / \Lsun)~\leq~-5.86$ \citep[by inferring
radii using the relations in ][]{burrows2001} and thus a
distance of 8--12~pc. 
Table~\ref{tab:props} summarises the inferred properties for the
  T~dwarfs identified in this paper.

 \begin{table*}
\centering
\begin{tabular}{|c c c c c c c|}
  \hline
Object & SpType & $T_{\rm eff}$ & log~{\it g} & mass & age & distance \\
       &  & K & log(cm s$_{-2}$) & M$_J$ & Gyr & pc \\
\hline
ULAS1017 & T8p & 750--850  & 5.0--5.5 & 33--70 & 1.6--15 & 31--54 \\
ULAS1335 & T9 & 550--600 & $<$5.0 & 15--31 & 0.6--5.3 & 8--12 \\
\hline
\end{tabular} 
\caption{A summary of our estimates for the properties of ULAS1017 and
  ULAS1335. Estimates for ULAS1238 have been generally neglected in 
 the absence of full flux calibrated spectral coverage. Distances
  assume that the dwarfs are single objects.
\label{tab:props}
}
 
\end{table*}

\section{Summary}
\label{sec:summ}

We have identified three very late-type T~dwarfs in the UKIDSS LAS
DR3: ULAS1017, ULAS1238 and ULAS1335, for which we have adopted the
spectral types T8p, T8.5 and T9 respectively.
ULAS1017 has a peculiar spectrum, with a $J$-band typical of a T8
dwarf, but $H$~and $K$-band peaks of a T6. 
In assigning spectral types to the our targets, we have defined the
extension of the T-dwarf sequence to the type T9, using the
H$_2$O--$H$ and $W_J$ indices, and ULAS1335 as the spectral standard.

To estimate atmospheric parameters for ULAS1017 and ULAS1335, we have
performed a detailed comparison with BT-Settl model spectra.
For ULAS1017, we have estimated that its temperature lies in the range
$750 \leq T_{\rm eff} \leq 850$, with $\log {\it g} = 5.0-5.5$
(assuming solar metallicity).

Our model comparison for ULAS1335 has suggested a $T_{\rm eff}$ that
is cooler than the coolest available BT-Settl models ($T_{\rm eff} < 
600{\rm K}$), and it may have low surface gravity possibly combined
with high metallicity.
As a result we have based our ``best-guess'' of the
atmospheric parameters on an examination of the near- to mid-infrared
SED. 
We have found ULAS1335 to be the reddest T-dwarf yet identified, with
$H-{\rm [4.49]}~=~4.34 \pm 0.04$. 
By extrapolating the empirical $T_{\rm eff}$ -- $H$--[4.49] relation
to lower $T_{\rm eff}$, and guided by model trends, we have estimated
that ULAS1335 has $T_{\rm eff} \sim 550-600{\rm K}$.
This estimate suggests that ULAS1335 is the coolest brown dwarf yet
discovered, although we await the determination of its distance by
parallax with interest.

\section*{Acknowledgments}

We wish to thank our anonymous referee for helpful suggestions that
have improved the quality of this paper.
We also thank P. Delorme for supplying us with an electronic version of the
spectrum for CFBDS0059 for use in this work.
We are extremely grateful to D. Saumon and M. Marley for providing us
with model $H$-[4.49] colours for their latest set of ultracool dwarf
atmospheres.
The UKIDSS project is defined in \citet{ukidss}.
UKIDSS uses the
UKIRT Wide Field Camera \citep[WFCAM;][]{wfcam} and a photometric
system described in \citet{hewett06}. The pipeline processing and
science archive are described in Irwin et al (2008) and \citet{WSA}. 
We have used data from the 3rd data release, which is
described in detail in Warren et al. (in prep). 
Based on observations made with the European Southern Observatory
telescopes obtained from the ESO/ST-ECF Science Archive Facility.
Results reported here are based on observations obtained at the Gemini
Observatory  under program numbers GN-2007B-Q-26 and
GN-2008A-Q-15. 
Gemini Observatory is operated by the Association of Universities for
Research in Astronomy, Inc. (AURA), under a cooperative agreement with
the NSF on behalf of the Gemini partnership: 
the National Science Foundation (United States),
the Science and Technology Facilities Council (United Kingdom), the
National Research Council (Canada), CONICYT (Chile), the Australian
Research Council (Australia), Ministério da Ciência e Tecnologia
(Brazil) and SECYT (Argentina).
SKL is supported by the Gemini Observatory, which is operated by AURA,
on behalf of the international Gemini partnership of Argentina,
Australia, Brazil, Canada, Chile, the United Kingdom, and the United
States of America. 
DBN is supported by Ministerio de Ciencia e Innovaci\'on (Spain).
This research has made use of the SIMBAD database,
operated at CDS, Strasbourg, France.
Research has benefited from the M, L, and T dwarf compendium housed
at DwarfArchives.org and maintained by Chris Gelino, Davy Kirkpatrick,
and Adam Burgasser. 
This research has benefited from the SpeX Prism Spectral Libraries,
maintained by Adam Burgasser at http://www.browndwarfs.org/spexprism.

\bibliographystyle{mn2e}
\bibliography{refs}

\end{document}